\documentclass[a4paper,11pt]{article}
\usepackage{amsmath,amsfonts,amsthm,color,psfrag}
\usepackage{pstricks,pst-plot}
\usepackage{enumerate}
\SpecialCoor

\newtheorem{theorem}{Theorem}
\newtheorem{lemma}[theorem]{Lemma}
\newtheorem{corollary}[theorem]{Corollary}

\def\Ghat{\widehat G}
\def\hatG{\widehat G}
\def\Ehat{\widehat E}
\def\Vhat{\widehat V}
\def\hatQ{\widehat Q}
\def\Khat{\widehat K}
\def\calA{\mathcal{A}}
\def\qbar{\bar q}
\def\abar{\bar a}
\def\wmod{\widetilde w}
\def\Bbar{\overline B}
\def\subsum{\mathit{\Sigma}}
\def\poly{\mathop{\mathrm{poly}}}
\def\NP{\mathrm{NP}}
\def\RP{\mathrm{RP}}
\def\numP{\mathrm{\#P}}
\def\numPQ{\numP_{\mathbb{Q}}}
\def\GapP{\mathrm{GapP}}

\def\tutte{\textsc{Tutte}}
\def\multitutte{\textsc{MultiTutte}}
\def\zerotutte{\textsc{ZeroMultiTutte}}
\def\PM{\#\textsc{Perfect Matchings}}
\def\APred{\leq_\mathrm{AP}}
\def\APequiv{\equiv_\mathrm{AP}}

\def\alphabet{\Sigma}
\def\APeq{\equiv_\mathrm{AP}}
\def\SAT{\textsc{\#Sat}}

\let\epsilon=\varepsilon
\let\rho=\varrho
\makeatletter
\def\prob#1#2#3{\goodbreak\begin{list}{}{\labelwidth\z@ \itemindent-\leftmargin
                        \itemsep\z@  \topsep6\p@\@plus6\p@
                        \let\makelabel\descriptionlabel}
                \item[\it Name.]#1
                \item[\it Instance.]#2
                \item[\it Output.]#3
                \end{list}}
\makeatother

\begin{document}
\title{Inapproximability of the Tutte polynomial\thanks{
Partially supported by the EPSRC grant Discontinuous Behaviour in the
Complexity of Randomized Algorithms. A preliminary version
of this paper appeared in the proceedings of the ACM {\it Symposium on
Theory of Computing\/}~\cite{STOCversion}}}
\author{Leslie Ann Goldberg\\
Department of Computer Science\\
University of Liverpool\\
Ashton Bldg, Liverpool L69 3BX, UK \and
Mark Jerrum \\
School of Mathematical Sciences\\
Queen Mary, University of London\\
London E1 4NS, UK }

\maketitle

\begin{abstract}
The Tutte polynomial of a graph~$G$
is
a two-variable polynomial $T(G;x,y)$
that encodes many interesting properties of the graph.
We study the complexity of the following problem,
for rationals~$x$ and~$y$:
take as input a graph $G$, and output
a value which is a good approximation to $T(G;x,y)$.
Jaeger, Vertigan and Welsh
have completely mapped the complexity of {\it exactly\/} computing
the Tutte polynomial. They have shown that this is \#P-hard,
except along the hyperbola $(x-1)(y-1)=1$
and at four special points.
We are interested in determining for which points $(x,y)$
there is a {\it fully polynomial randomised approximation scheme\/}
(FPRAS)
for $T(G;x,y)$.
Under the assumption $\RP\neq \NP$, we prove that
there is no FPRAS at $(x,y)$ if $(x,y)$ is is in one of the half-planes
$x<-1$ or $y<-1$ (excluding the easy-to-compute cases mentioned
above).
Two exceptions to this result are the half-line $x<-1,y=1$
(which is still open) and the portion of
the hyperbola $(x-1)(y-1)=2$
corresponding to $y<-1$
which we show to be  equivalent in difficulty to approximately counting
perfect matchings.
We give further intractability results for $(x,y)$ in the vicinity of
the origin.
A corollary of our results is that,
under the assumption $\RP\neq\NP$, there is no
FPRAS at the point $(x,y)=(0,1-\lambda)$ when $\lambda>2$ is a positive
integer. Thus, there is no FPRAS for counting nowhere-zero $\lambda$ flows
for $\lambda>2$. This is an interesting consequence of our work
since the corresponding decision
problem is in P for example for $\lambda=6$.
Although our main concern is to distinguish regions of the Tutte plane
that admit an FPRAS from those that do not, we also note that the
latter regions exhibit different levels of intractability.
At certain points $(x,y)$, for example the integer points on the
$x$-axis, or any point in the positive quadrant,
there is a randomised approximation scheme for $T(G;x,y)$ that
runs in polynomial time using an oracle for an NP predicate.
On the other hand,
we identify
a region of points $(x,y)$
at which even approximating $T(G;x,y)$
is as hard as~\#P.

\end{abstract}

\section{Summary of results}
\label{sec:summary}

The Tutte polynomial of a graph~$G$
(see Section~\ref{sec:tuttedef})
is
a two-variable polynomial $T(G;x,y)$
that encodes many interesting properties of the graph.
We mention only some of these properties here, as
a longer and more detailed list can be found
in Welsh's book~\cite{Welsh93}.
\begin{itemize}
\item $T(G;1,1)$ counts the number of spanning trees
of a connected graph  $G$.
\item $T(G;2,1)$ counts the number of forests
in~$G$ (the number of edge subsets that contain no cycles).
\item $T(G;1,2)$ counts the number of edge subsets that are connected
and span $G$.
\item $T(G;2,0)$ counts the number of acyclic orientations of $G$.
\item The chromatic polynomial
$P(G;\lambda)$ of a graph~$G$ with $n$ vertices, $m$ edges and
$k$ connected components is given
by
$$P(G;\lambda)= {(-1)}^{n-k} \lambda^{k}
T(G;1-\lambda,0).$$
When $\lambda$ is a positive integer, $P(G;\lambda)$ counts
the proper $\lambda$-colourings of~$G$.
\item The flow polynomial $F(G;\lambda)$ is given
by
$$F(G;\lambda) = {(-1)}^{m -n+k} T(G;0,1-\lambda).$$
When $\lambda$ is a positive integer, $F(G;\lambda)$
counts the nowhere-zero $\lambda$-flows of~$G$.
\item The all-terminal reliability polynomial
$R(G;p)$ is given by
$$R(G;p) = {(1-p)}^{m - n+k}p^{n-k} T(G;1,1/(1-p)).$$
When $G$ is connected and
each edge is independently ``open''
with probability~$p$,
$R(G,p)$
is the probability that there is a path between every pair
of vertices of~$G$.
\item For every positive integer~$q$,
the Tutte polynomial
along the hyperbola $H_q$ given by $(x-1)(y-1)=q$ corresponds to
the partition function of the $q$-state Potts model.
\end{itemize}

We study the complexity of the following problem,
for rationals~$x$ and~$y$:
take as input a graph $G$, and output
a value which is a good approximation to $T(G;x,y)$.
Jaeger, Vertigan and Welsh~\cite{JVW90}
(see Section~\ref{sec:prev})
have completely mapped the complexity of {\it exactly\/} computing
the Tutte polynomial. They have shown that this is \#P-hard,
except along the hyperbola $H_1$
and at the four special points $(x,y)\in\{(1,1),(0,-1),(-1,0),(-1,-1)\}$.
(\#P is the analogue, for counting problems, of the more familiar
class NP of decision problems.)

We are interested in determining for which points $(x,y)$
there is a {\it fully polynomial randomised approximation scheme\/}
(FPRAS)
for $T(G;x,y)$.
An FPRAS is a polynomial-time randomised approximation algorithm
achieving arbitrarily small relative error.
Precise definitions of FPRAS, \#P, and other
complexity-theoretic terminology will be provided in
Section~\ref{sec:fpras}.

It is known that there is an FPRAS for every point $(x,y)$
on the hyperbola $H_2$ with $y>1$ --- this follows
from the Ising result of Jerrum and Sinclair~\cite{JS93}.
No other general FPRAS results are known.
A few negative results are known --- see Section~\ref{sec:prev}.

Our
goal is to map the Tutte plane in terms of FPRASability as completely
as possible.  The specific contribution of this article is a substantial
widening of the region known to be non-FPRASable.

\begin{figure}[h]
\centering

\def\N{8}%the max value of x (and of y)

\psset{unit=0.6}

 \begin{pspicture}(-\N,-\N)(\N,\N)
  \psgrid[griddots=10,gridlabels=0pt, subgriddiv=0]
  \psaxes(0,0)(\N,\N)(\N,\N)

%The hard half-plane x<-1
\pspolygon[fillcolor=gray,fillstyle=solid](-8,-8)(-1,-8)(-1,8)(-8,8)
%re-draw x=-1 in white because we don't know about the line
%H_1 will get filled in later
\psline[linewidth=3pt,linecolor=white](-1,8)(-1,-8)
\psset{framesep=1.5pt}
\rput(-4.7,7){\psframebox*{$x<-1$ except $q=0,1$}}

%The hard half-plane y<-1
\pspolygon[fillcolor=gray,fillstyle=solid,linestyle=none](-8,-8)(-8,-1)(8,-1)(8,-8)
%re-draw y=-1 in white because we don't know about the line
%H_1 and H_2 will get filled in later
\psline[linewidth=3pt,linecolor=white](8,-1)(-1,-1)
\psset{framesep=1.5pt}
\rput(-4.7,-7){\psframebox*{$y<-1$ except $q=1,2$}}

%H0
%\psline[linewidth=2pt,linecolor=white](1,-1)(1,-8)
%%commented the above one out because we can now do
%%x=1,y<-1
%stop at -1 to avoid overwriting
\psline[linewidth=2pt,linecolor=white](-1,1)(-8,1)

%%H2
%%x-1=t y-1=(2/t)
\parametricplot[linecolor=red,linewidth=2pt]%
{-1}{-2 9 div}{1 t add 1 2 t div add}
%%min value of t gives x=0 (y=-1) so is t=-1 (stop at y=-1 because rest is
%in previous region)
%%max value of t gives y=-8 so is t=-2/9

%ferro ISING
%I'm going to colour all FPRAS green for now so that we have
% more colours left to distinguish between other things
\parametricplot[linecolor=green,linewidth=2pt]%
{2 7 div}{7}{1 t add 1 2 t div add}
% q=2

%Ferro 3-Potts  (no need to draw since we know nothing)
%   \parametricplot[linecolor=blue,linewidth=2pt]%
%  {3 7 div}{7}{1 t add 1 3 t div add}

%triangles
\pspolygon[fillcolor=gray,fillstyle=solid,linecolor=white](-1,1)(-1,-1)(0,-1)
%lemma 7
\pspolygon[fillcolor=gray,fillstyle=solid,linecolor=white](1,-1)(-1,-1)(-1,0)
%lemma 8
%I have coloured boundaries white, but the boundary of the one triangle
%that is in the interior of the other really is known to be hard:
\psline[linewidth=3pt,linecolor=gray](-1,0)(-0.333,-0.333)
%And a little bit of the boundary can be done
\psline[linewidth=3pt,linecolor=gray](-1,-1)(-1,0.29)
\psline[linewidth=3pt,linecolor=gray](-1,-1)(0.29,-1)

%An extra hard region.
%Here we are working within |x|<1 and |y|<1 and q<=1.5
   \pscustom[linewidth=1.5pt,linecolor=gray]{%
\parametricplot[linecolor=gray,linewidth=2pt]%
  {-2}{-1.5 2 div}{1 t add 1 1.5 t div add}
  \gsave
  \psline(-0.9,-0.9)
  \fill[fillstyle=solid,fillcolor=gray,linecolor=gray]
  \grestore
  }

%H1
%x-1=t
%y-1=(1/t)
 \parametricplot[linecolor=green,linewidth=2pt]%
  {1 7 div}{7}{1 t add 1 1 t div add}
%max value of t gives x=8 so is t=7
%The "8" is \N, but it seems we can't use macros here
%min value of t gives y=8 so is t=1/7
\parametricplot[linecolor=green,linewidth=2pt]%
  {-9}{-1 9 div}{1 t add 1 1 t div add}
%min value of t gives x=-8 so is t=-9
%max value of t gives y=-8 so is t=-1/9

%%H4
%%x-1=t y-1=(4/t)
\parametricplot[linecolor=black,linewidth=2pt]%
{-4}{-2}{1 t add 1 4 t div add}

%special points

\psset{linecolor=green}
\qdisk(1,1){4pt}
\qdisk(-1,0){4pt}
\qdisk(-1,-1){4pt}
\qdisk(0,-1){4pt}

\end{pspicture}

\caption{Green points are FPRASable, red points are equivalent to perfect matchings
and gray points are not FPRASable unless RP=NP.
We don't know about white points.
The points depicted in black are
at least as hard as gray and are presumably harder  --- this is the region of $q=4$
with $y\in(-1,0)$ and approximating Tutte is actually \#P-hard here.
(There are presumably more such points.)}

\label{fig:tuttepic}
\end{figure}
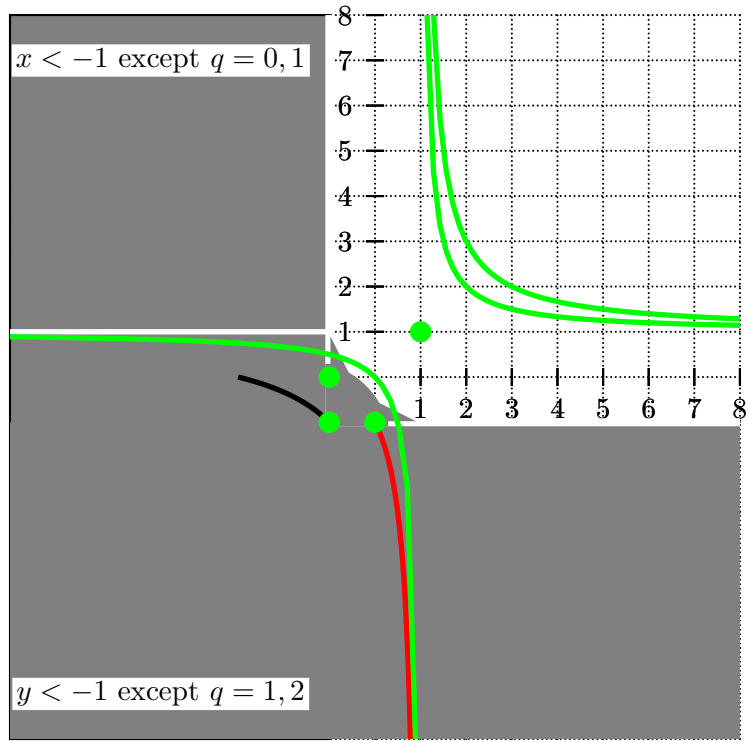

Our contributions are summarised in
Figure~\ref{fig:tuttepic}.
In particular, under the assumption $\RP\neq \NP$, we prove the following.
\begin{enumerate}[(1)]
\item If $x<-1$ and $(x,y)$ is not on $H_0$ or $H_1$, then
there is no FPRAS at $(x,y)$ (Corollary~\ref{cor:x<-1}).
\item If $y<-1$ and $(x,y)$ is not on $H_1$ or $H_2$,
then there is no FPRAS at $(x,y)$ (Corollary~\ref{cor:y<-1}
when $(x,y)$ is not on
 $H_0$
and Lemma~\ref{lem:x=1} for the case in which $(x,y)$ is on $H_0$).
\item If $(x,y)$ is on $H_2$ and $y<-1$ then approximating
$T(G;x,y)$ is equivalent in difficulty to approximately counting
perfect matchings (Lemma~\ref{lem:matchings}).
\item If $(x,y)$ is not on $H_1$ and
is in the vicinity of the origin in the sense that $|x|<1$ and
$|y|<1$ and is in the triangle
given by $y<-1-2x$ then there is no FPRAS (Lemma~\ref{lem:triangle1}).
\item If $(x,y)$ is not on $H_1$ and is in the vicinity of the
origin and is in the triangle given by
$x<-1-2y$ then there is no FPRAS
(Lemma~\ref{lem:triangle2}).
\item The two previous intractability results
(results (4) and (5))
can be partially extended to the boundary of the triangles
(Lemma~\ref{lem:boundary1} and \ref{lem:boundary2}).
\item If $(x,y)$ is in the vicinity of the origin
and $q=(x-1)(y-1)>1.5$ then there is no FPRAS (excluding
the special points at which exact computation is possible)
(Lemma~\ref{lem:vicinity}).
\end{enumerate}

Result~(2) above implies that, under the assumption $\RP\neq\NP$, there is no
FPRAS at the point $(x,y)=(0,1-\lambda)$ when $\lambda>2$ is a positive
integer. Thus, there is no FPRAS for counting nowhere-zero $\lambda$ flows
for $\lambda>2$. This is an interesting consequence of our work
since Seymour~\cite{seymour} has shown that the corresponding decision
problem is in P for $\lambda=6$. In particular,
a graph has a $6$-flow if and only if it has no bridge (cut edge).

Although our main concern is to distinguish regions of the Tutte plane
that admit an FPRAS from those that do not, we also note that the
latter regions exhibit different levels of intractability.
At certain points $(x,y)$, for example the integer points on the
$x$-axis, or any point in the positive quadrant,
there is a randomised approximation scheme for $\tutte(x,y)$ that
runs in polynomial time using an oracle for an NP predicate.
On the other hand,
Theorem~\ref{thm:nump}
identifies a region of points $(x,y)$
at which even approximating $\tutte(x,y)$
is as hard as~\#P.  These two kinds of intractability are
very different, assuming \#P is a ``much bigger'' class than
NP\null.

\section{Definitions and context}

\subsection{The Tutte polynomial}
\label{sec:tuttedef}

The Tutte polynomial of a graph $G=(V,E)$
is the two-variable polynomial

\begin{equation}
\label{eq:tutte}
T(G;x,y) = \sum_{A\subseteq E} {(x-1)}^{\kappa(A)-\kappa(E)}
{(y-1)}^{|A|-n+\kappa(A)},
\end{equation}
where $\kappa(A)$ denotes the number of connected components of the graph $(V,A)$
and $n=|V|$.
Following the usual convention for the Tutte polynomial~\cite{Sokal05} a graph is allowed
to have loops and/or multiple edges, and we use the term ``graph'' in this way except where we
explicitly state otherwise. The Tutte
polynomial is sometimes referred to as the ``Whitney-Tutte'' polynomial,
or the ``dichromatic polynomial''.
See \cite{Tutte84} and \cite{Welsh93}.

\subsection{The complexity of
counting and approximate counting} \label{sec:fpras}

We start with a brief summary of
the complexity of counting. See \cite{Jerrum} for more details.
A counting problem can be viewed as a function $f:\Sigma^* \rightarrow
\mathbb{N}$ mapping an encoding of a problem instance
(encoded as a word in a finite alphabet, $\Sigma$)
to a natural number.
For example, $f$
might map an encoding of a graph~$G$ to the number of independent
sets of~$G$.
\#P is the analogue of NP for counting problems.
A counting problem $f:\Sigma^*\rightarrow\mathbb{N}$ is in \#P
if there is a polynomial-time predicate $\chi: \Sigma^* \times
\Sigma^*\rightarrow \{0,1\}$ and a polynomial $p$ such that, for
all instances $x\in \Sigma^*$,
$$f(x) = |\{w \in \Sigma^* \mid \chi(x,w) \wedge
|w| \leq p(|x|)
\}|.$$

It is straightforward to check that $\tutte(x,y)\in\numP$ when
$x,y$ are integers with $x,y\geq1$.  If $x,y$ are arbitrary integers
then the terms in the Tutte polynomial vary in sign, and
the problem $\tutte(x,y)$ no longer fits the \#P framework.
In that case, $\tutte(x,y)\in\GapP$, where
$\GapP$ is the set of functions $f=f^{+}-f^{-}:\Sigma^{*}\to\mathbb{Z}$
expressible
as the difference of two \#P-functions $f^{+}$ and~$f^{-}$.
(Simply partition the
terms of the Tutte polynomial according to sign, and
compute the positive and negative parts separately.)

Finally, since we do not want to restrict ourselves to integer $x$ and~$y$,
we need to extend the classes \#P and $\GapP$ a little to
encompass computations over the rationals.  We say that
$f:\Sigma^{*}\to\mathbb{Q}$ is in the class $\numPQ$ if
$f(x)=a(x)/b(x)$, where $a,b:\Sigma^{*}\to\mathbb{N}$, and
$a\in\numP$ and $b\in\mathrm{FP}$, where  FP is  the class of
functions computable by polynomial-time algorithms.
If $x,y\geq 1$, then $\tutte(x,y)\in\numPQ$, since we may multiply
through by suitable powers of the denominators of $x$ and~$y$,
after which all the terms in the Tutte polynomial become integers.

A \emph{randomised approximation scheme\/} is an algorithm for
approximately computing the value of a function~$f:\Sigma^*\rightarrow
\mathbb{R}$. The
approximation scheme has a parameter~$\varepsilon>0$ which specifies
the error tolerance.
A \emph{randomised approximation scheme\/} for~$f$ is a
randomised algorithm that takes as input an instance $ x\in
\alphabet^{\ast }$ (e.g., an encoding of a graph~$G$) and an error
tolerance $\varepsilon >0$, and outputs a number $z\in \mathbb{Q}$
(a random variable of the ``coin tosses'' made by the algorithm)
such that, for every instance~$x$,
\begin{equation}
\label{eq:3:FPRASerrorprob}
\Pr \big[e^{-\epsilon} f(x)\leq z \leq e^\epsilon f(x)\big]\geq \frac{3}{4}\, .
\end{equation}
The randomised approximation scheme is said to be a
\emph{fully polynomial randomised approximation scheme},
or \emph{FPRAS},
if it runs in time bounded by a polynomial
in $ |x| $ and $ \epsilon^{-1} $.
Note that the quantity $3/4$ in
Equation~(\ref{eq:3:FPRASerrorprob})
could be changed to any value in the open
interval $(\frac12,1)$ without changing the set of problems
that have randomised approximation schemes~\cite[Lemma~6.1]{JVV86}.

It is known that every counting problem in~\#P
has a randomised approximation scheme
whose complexity is not much greater than NP\null.
In particular, if $f$ is a counting problem in~\#P
then the bisection technique of Valiant and Vazirani
\cite[Cor 3.6]{vv} can be used to construct
a randomised approximation scheme for $f$ that
runs in polynomial time, using an oracle for
an NP~predicate.  See \cite[Theorem 3.4]{JVV86}
or \cite[Theorem 1]{BIS};  also \cite{Stock83} for
an early result in this direction.

We will use
the notion of approximation-preserving reduction
from Dyer, Goldberg, Greenhill and Jerrum~\cite{BIS}. Suppose that $f$ and $g$ are functions from
$\alphabet^{\ast }$ to~$\mathbb{R}$. An ``approximation-preserving
reduction'' from~$f$ to~$g$ gives a way to turn an FPRAS for~$g$
into an FPRAS for~$f$. An {\it approximation-preserving reduction\/}
from $f$ to~$g$ is a randomised algorithm~$\mathcal{A}$ for
computing~$f$ using an oracle for~$g$. The algorithm~$\mathcal{A}$ takes
as input a pair $(x,\varepsilon)\in\alphabet^*\times(0,1)$, and
satisfies the following three conditions: (i)~every oracle call made
by~$\mathcal{A}$ is of the form $(w,\delta)$, where
$w\in\alphabet^*$ is an instance of~$g$, and $0<\delta<1$ is an
error bound satisfying $\delta^{-1}\leq\poly(|x|,
\varepsilon^{-1})$; (ii) the algorithm~$\mathcal{A}$ meets the
specification for being a randomised approximation scheme for~$f$
(as described above) whenever the oracle meets the specification for
being a randomised approximation scheme for~$g$; and (iii)~the
run-time of~$\mathcal{A}$ is polynomial in $|x|$ and
$\varepsilon^{-1}$.

If an approximation-preserving reduction from $f$ to~$g$
exists we write $f\APred g$, and say that {\it $f$ is AP-reducible to~$g$}.
Note that if $f\APred g$ and $g$ has an FPRAS then $f$ has an FPRAS\null.
(The definition of AP-reduction was chosen to make this true).
If $f\APred g$ and $g\APred f$ then we say that
{\it $f$ and $g$ are AP-interreducible}, and write $f\APeq g$.

Dyer et al.~\cite{BIS}
identified three classes of counting problems that are interreducible
under approximation-preserving reductions. The first class, containing the
problems that admit an FPRAS, are trivially AP-interreducible since
all the work can be embedded into the reduction (which declines to
use the oracle). The second class (and the last one
that we will describe here) is the set of problems that are
AP-interreducible with \SAT, the problem of counting
satisfying assignments to a Boolean formula in CNF\null.
Zuckerman~\cite{zuckerman}
has shown that \SAT{} cannot have an FPRAS unless
$\mathrm{RP}=\mathrm{NP}$. The same is obviously true of any problem
in $\numP$ to which \SAT{} is AP-reducible. See~\cite{BIS} for
details.

\subsection{The Tutte polynomial and \#P}
\label{sec:innumP}

We will study the following computational problem for
rationals $x$ and $y$.
\begin{description}
\item[Name] $\tutte(x,y)$.
    \item[Instance]  A graph $G=(V,E)$.
    \item[Output]  $T(G;x,y)$.
\end{description}

Given fixed rationals $x$ and $y$, $\tutte(x,y)$
is a function from $\Sigma^*$ to $\mathbb{Q}$,
mapping an encoding of a graph $G$ to a rational
$T(G;x,y)$.
It is not immediately clear from the definition
(\ref{eq:tutte})
that $\tutte(x,y)$ is in
$\numPQ$, but this is known to be true
if $x$ and $y$ are both non-negative.

In particular, if $G$ is connected, it
is known~\cite{Tutte54} (see also
\cite[Theorem 1X.65]{Tutte84}) that
$T(G;x,y)$ can be expressed as
$T(G;x,y) = \sum_{T} x^{a(T)} y^{b(T)},$
where the sum is over spanning trees~$T$ of~$G$, and
$a(T)$ and $b(T)$ are
natural numbers which are
easily computable from~$T$.\footnote{Indeed, historically, this appears to
have been the original definition of the polynomial.}
($a(T)$ is the number of so-called ``internally active'' edges of~$T$ and
$b(T)$ is the number of ``externally active'' edges of~$T$ ---
see \cite{Tutte84} for details.)

It is clear from the definition~(\ref{eq:tutte}) that the
Tutte polynomial of a
graph $G$ (which may have several connected components)
may be expressed as a product of the Tutte polynomials
of the components.
Thus, for $x\geq 0$ and $y\geq 0$,
we have $\tutte(x,y) \in\numPQ$, which means
that there is a randomised approximation scheme for
$\tutte(x,y)$ that runs in polynomial time, using an oracle
for an NP~predicate.

It is unlikely that $\tutte(x,y)$
is in $\numPQ$ for
all $x$ and $y$.
In particular, Theorem~\ref{thm:nump} identifies
a region of points $(x,y)$, where $y$ is negative, for which
even approximating $\tutte(x,y)$ is as hard as \#P.

\subsection{Previous work on the complexity of
the Tutte polynomial}
\label{sec:prev}

Jaeger, Vertigan and Welsh~\cite{JVW90} have
completely mapped the complexity of
exactly computing the Tutte polynomial.
They  have observed that $\tutte(x,y)$ is
in FP for any point $(x,y)$ on the hyperbola $H_1$.
This can be seen from the definition (\ref{eq:tutte}), since
terms involving $\kappa(A)$ cancel.
Also, $\tutte(x,y)$ is in FP when $(x,y)$
is one of the special points
$(1,1)$, $(0,-1)$, $(-1,0)$, and $(-1,-1)$.
As noted in Section~\ref{sec:summary},
$T(G;1,1)$ is the number of spanning trees of a connected graph~$G$,
$T(G;0,-1)$ is the number of $2$-flows of $G$
(up to a factor of plus or minus one), and
$T(G;-1,0)$ is the number of $2$-colourings of $G$
(up to an easily computable factor).
$T(G;-1,-1)$ has an interpretation in terms of the ``bicycle
space'' of~$G$. See \cite[(2.8)]{JVW90}.
Intriguingly, Jaeger, Vertigan and Welsh managed to
show that $\tutte(x,y)$ is \#P-hard for every other
pair of rationals $(x,y)$.
They also investigated
the complexity of
evaluating the Tutte polynomial when $x$ and $y$
are real or complex numbers, but that is beyond the scope of this paper.

The only FPRAS   for approximating the
Tutte polynomial that we know of is the
Ferromagnetic Ising FPRAS of Jerrum and Sinclair~\cite{JS93}.
This gives an FPRAS for $\tutte(x,y)$ for
every point $(x,y)$ on $H_2$ with  $y>1$.
The connection between the Ising model and the Tutte
polynomial along the hyperbola~$H_2$ is elaborated
later in the paper --- see Equation~(\ref{eq:Potts}).
We know of no other FPRASes for approximating the Tutte polynomial
for an arbitrary input graph~$G$.
There is some related work, however, for example,
Karger~\cite{karger} gives an FPRAS for {\bf non}-reliability, which
is not the same thing as an FPRAS for reliability,
but is somewhat related.

There are also FPRASes known for special cases in which
restrictions are placed on~$G$.
For example, \cite{afw} gives an FPRAS for
points $(x,y)$ with $x\geq 1$ and $y\geq 1$
for the restricted case in which the input graph $G$
is ``dense'', meaning that the $n$-vertex graph $G$ has
minimum degree $\Omega(n)$.
As another example, there is a huge literature
on approximately counting proper colourings of degree-bounded
graphs.

Several negative results are known
for  approximating the Tutte polynomial.
First, note that if $T(G;x,y)$ is the number of solutions to an
NP-complete decision problem, then there can be no FPRAS
for $\tutte(x,y)$ unless $\RP=\NP$.
So, for example, if $\lambda>2$ is a positive integer,
then by the chromatic polynomial specialisation mentioned in
Section~\ref{sec:summary}, there is no FPRAS for
$T(G;1-\lambda,0)$.

Jerrum and Sinclair~\cite[Theorem 14]{JS93} showed that there is
no FPRAS for (antiferromagnetic) Ising unless $\RP=\NP$.
This implies that, unless $\RP=\NP$, there is no FPRAS for the
function whose input is a graph $G$ and
a point $(x,y)$ on $H_2$ with  $0<y<1$
and whose output is $T(G;x,y)$.

Welsh~\cite{Welsh94} extended this result. Specifically, he showed
the following,
assuming $\RP\neq\NP$.
\begin{itemize}
\item Suppose $q\geq 2$ is a positive integer.
Then there is no FPRAS for the function whose input is a graph $G$
and a point $(x,y)$ on $H_q$ with $x<0,y>0$ and
whose output is $T(G;x,y)$.
Furthermore, there is no FPRAS for the function whose input
is a graph $G$ and a point $(x,y)$ on $H_q$ with
$x<0,y<0$ and whose output is $T(G;x,y)$.
\item There is no FPRAS for the function whose input is a graph
$G$ and a point $(x,y)$ on $H_3$ with  $0<x<1$
and whose output is $T(G;x,y)$.
\end{itemize}

\section{Regions of the Tutte plane that do not admit an
FPRAS unless RP=NP}\label{sec:map}

The tensor product of matroids was introduced by
Brylawski~\cite{Bry82}.
We define it here in the special case of graphs.
Let $G$ be a graph, and $K$
another graph with a distinguished
edge $e$ with endpoints $u$ and~$u'$.
The tensor product $G\otimes K$ is obtained
from~$G$ by performing a {\it $2$-sum\/} operation with~$K$ on each
edge~$f$ of~$G$ in turn:  Let the endpoints of~$f$ be $v$ and~$v'$.
Take a copy of~$K$ and identify vertex $u$ (resp.~$u'$)
of~$K$ with $v$ (resp.~$v'$) of~$G$, and then delete edges
$e$ and~$f$.  (Since $G$ and $K$ are undirected graphs,
there are two ways of performing the $2$-sum.  This lack of
uniqueness is an artefact of viewing a matroid operation in
terms of graphs, which have additional structure.  However,
the Tutte polynomial is insensitive to which of the two possible
identifications is made.)
For technical reasons we will assume that $e$ is not a bridge
of $K$. In particular, we assume that
deleting $e$ does not increase the number of connected components of~$K$.

Let $K\setminus e$ be the graph constructed
from $K$ by deleting edge~$e$. Let $K/e$ be the graph constructed
from $K$ by contracting edge~$e$.
Suppose $(x,y)\in\mathbb{Q}^2$.
Let $q=(x-1)(y-1)$.
Define the point $(x',y')$ as follows.

\begin{align}
    x'&=\frac{(1-q)T(K\setminus e;x,y)}{T(K\setminus e;x,y)-
    (x-1)\,T(K/e;x,y)}
    \label{eq:x'defn}\\
\noalign{\noindent and}
    y'&=\frac{(1-q)T(K/e;x,y)}{T(K/e;x,y)-
    (y-1)T(K\setminus e;x,y)}.
    \label{eq:y'defn}
\end{align}
Then it is known
(\cite[(4.1)]{JVW90})
that

\begin{equation}\label{eq:tensor}
T(G;x',y') =
L(x,y,K)^{m}M(x,y,K)^{n-\kappa}
T(G\otimes K;x,y),
\end{equation}
where
$n$, $m$ and $\kappa$ are (respectively) the number of
vertices, edges and connected components in~$G$
and
$$
L(x,y,K)=
\frac{1-q}{
T(K/e;x,y)-
    (y-1)T(K\setminus e;x,y)
},$$
and
$$M(x,y,K)=
\frac{T(K/e;x,y)-
    (y-1)T(K\setminus e;x,y)}
    {T(K\setminus e;x,y)-
    (x-1)\,T(K/e;x,y)}
 .$$

Suppose that the denominators of (\ref{eq:x'defn}) and (\ref{eq:y'defn}) are non-zero.
In this case, the point $(x',y')$ is well-defined and we say that
$(x,y)$ is {\it shifted to\/} the point $(x',y')$ by~$K$.
In this case, $L(x,y,K)$ and $M(x,y,K)$ are also well-defined,
so Equation~(\ref{eq:tensor}) gives us the reduction
$\tutte(x',y')\APred  \tutte(x,y)$.

We will not prove Equation~(\ref{eq:tensor}) since the equation can be found
elsewhere (eg., \cite[(4.1)]{JVW90}) but, for completeness, we will derive
similar identities that we will use in Section~\ref{sec:reductions}.
We are particularly interested in
two special cases. The
case in which  $K$ is a
cycle on $k+1$ vertices is known as a {\it $k$-stretch\/}
in the literature and the case in which
$K$ is a two-vertex graph with $k+1$ parallel
edges is known as a {\it $k$-thickening\/}.  Informally, a $k$-stretch
of~$G$ replaces each edge of $G$ by a path of length~$k$, while
a $k$-thickening replaces each edge by a bundle of $k$ parallel
edges.
Specifically,
\begin{equation}
(x',y')=\begin{cases}
\big(x^k, q/(x^k-1)+1\big)&\text{for a $k$-stretch;}\\
\big(q/(y^k-1)+1, y^k\big)&\text{for a $k$-thickening.}
\end{cases}\label{eq:stretchThickenRules}
\end{equation}

Observe that $q=(x-1)(y-1)$ is an invariant
for stretches and thickenings, and indeed
for shifts in general.  It is this limitation that
gives the hyperbolas $H_q$ a special place in the complexity
theory of the Tutte polynomial.
All shifts preserve $q=(x-1)(y-1)$ but not all AP-reductions do. In particular,
the construction in \cite[(5.12)]{JVW90} (taking $p=1$),
based on an idea of Linial~\cite{linial},
gives the reduction $\tutte(x,0)\APred \tutte(x-1,0)$ for $x\neq 1$.

We shall make frequent use of the fact that shifts may be composed.

\begin{lemma}
The relation ``shifts to'' is transitive.
\end{lemma}

\begin{proof}
Suppose $K_{1}$ is a graph that implements the shift
$(x,y)\to(x',y')$ and $K_{2}$ is the graph (with distinguished
edge~$e$) that implements $(x',y')\to(x'',y'')$.
Let $\Khat$ be the graph obtained from $K_{2}$ by performing
a 2-sum with $K_{1}$ along every edge of $K_{2}$ except~$e$;
let $e$ remain the distinguished edge of~$\Khat$.  We claim that
$\Khat$ implements the shift $(x,y)\to (x'',y'')$.
Since $G\otimes\Khat=(G\otimes K_{2})\otimes K_{1}$,
for any~$G$, this ought to be true, but we can verify
the claim by direct calculation.

Evaluating the rhs of~(\ref{eq:x'defn}), with $K=\Khat$:
\begin{align}
&\frac{(1-q)T(\Khat\setminus e;x,y)}
  {T(\Khat\setminus e;x,y)-(x-1)T(\Khat/e;x,y)}\notag\\
&\qquad=\frac{(1-q)T((K_{2}\setminus e)\otimes K_{1};x,y)}
  {T((K_{2}\setminus e)\otimes K_{1});x,y)
   -(x-1)T((K_{2}/e)\otimes K_{1};x,y)}\notag\\
&\qquad=\frac{(1-q)T(K_{2}\setminus e;x',y')}
  {T(K_{2}\setminus e;x',y')
   -(x-1)M(x,y,K_{1})T(K_{2}/e;x',y')}\label{eq:use5}\\
&\qquad=\frac{(1-q)T(K_{2}\setminus e;x',y')}
  {T(K_{2}\setminus e;x',y')
   -(x'-1)T(K_{2}/e;x',y')}\label{eq:Mdef}\\
&\qquad=x''.\label{eq:use3}
\end{align}
Here, (\ref{eq:use3}) uses~(\ref{eq:x'defn}),
and (\ref{eq:Mdef}) the fact that $(x'-1)=(x-1)M(x,y,K_{1})$.
Equality~(\ref{eq:use5}) follows from~(\ref{eq:tensor}),
noting that $K_{2}/e$ has the same number of edges
as $K_{2}\setminus e$, but one fewer vertex.
A similar calculation holds for~$y''$.
\end{proof}

Shifts play a key role in the classical study of the
complexity of exact computation of the Tutte
polynomial~\cite{JVW90}, and the same is true in the
current investigation.  Our keys tools are the following.

\begin{theorem}\label{thm:shift}
Suppose $(x,y)\in \mathbb{Q}^2$ satisfies
$q=(x-1)(y-1)\notin\{0,1,2\}$.  Suppose also that
it is possible to shift the point
$(x,y)$ to the point $(x',y')$ with $y'\notin[-1,1]$,
and to $(x'',y'')$ with $y''\in(-1,1)$.
Then there is no FPRAS for
$\tutte(x,y)$
unless
$\mathrm{RP}=\mathrm{NP}$.
\end{theorem}

\begin{theorem}\label{thm:shift'}
Suppose $(x,y)\in \mathbb{Q}^2$ satisfies
$q=(x-1)(y-1)\notin\{0,1,2\}$.  Suppose also that
it is possible to shift the point
$(x,y)$ to the point $(x',y')$ with $x'\notin[-1,1]$,
and to $(x'',y'')$ with $x''\in(-1,1)$.
Then there is no FPRAS for
$\tutte(x,y)$
unless
$\mathrm{RP}=\mathrm{NP}$.
\end{theorem}

Since the notion of ``shift'' is defined for any class of
matroids closed under tensor product,
it should be possible to frame statements
similar to Theorems \ref{thm:shift} and~\ref{thm:shift'}
for classes of matroids other than graphic.
Although the two theorems are dual to one another in the
matroid theoretic sense, they are not equivalent,
since the class of graphic matroids is not closed
under duality.

The proofs of Theorem~\ref{thm:shift} and~\ref{thm:shift'} appear
in Section~\ref{sec:reductions}.

\subsection{Two halfplanes}
\begin{corollary}\label{cor:x<-1}
Suppose $(x,y)$ is a point lying in the open half-plane
$x<-1$ but not on the hyperbolas~$H_0$ or~$H_1$.
Under the assumption $\mathrm{RP}\not=\mathrm{NP}$
there is no FPRAS for $\tutte(x,y)$.
\end{corollary}
\begin{proof}
Let $(x,y)\in\mathbb{R}^2$ be a point not on $H_0$ or~$H_1$
that satisfies $x<-1$.  At the outset, we'll assume further
that $(x,y)\notin H_2$ and that $y\not=-1$. There are
three cases, depending on~$y$.
First assume $y>1$, and observe that $q=(x-1)(y-1)<0$.
Using a $k$-stretch, we may shift the point $(x,y)$ to the
the point $(x'',y'')= \big(x^k, q/(x^k-1)+1\big)$.  Now
$y''\in(-1,1)$ for all sufficiently large even~$k$
so Theorem~\ref{thm:shift} applies.  (The trivial shift,
taking $(x,y)$ to itself, provides the point $(x',y')$
with $y'\notin[-1,1]$.)
A similar argument, but setting $k$ to be large and odd
deals with the situation $y<-1$.
Finally, when $y\in(-1,1)$, a $2$-stretch shifts
$(x,y)$ to the point
$(x',y')=\big(x^2,q/(x^2-1)+1\big)=\big(x^2,(y-1)/(x+1)+1\big)$,
with $y'>1$.

The additional condition $y\not=-1$ may be removed
by noting that a $3$-stretch shifts $(x,-1)$
to a point $(x'',y'')=\big(x^3,1-2/(x^2+x+1)\big)$
with $x''<-1$ and $y''\in(-1,+1)$,
and we have already seen how to deal with such a point.

Finally, suppose $q=2$.
Like Welsh~\cite{Welsh94} we will show hardness using an
argument of Jerrum and Sinclair~\cite[(Theorem 14)]{JS93}.
Suppose
that $G$ has $n$ vertices and $m$ edges and that $x'$ and $y'$ satisfy
$(x'-1)(y'-1)=2$.
Jerrum and Sinclair note
that
$$T(G;x',y') = (y'-1)^n (x'-1)^{-\kappa(E)}
\sum_{r=0}^m N_r (y')^{m-r}$$
where $N_r$ is the number of functions $\sigma:V\rightarrow\{-1,1\}$
with $r$ bichromatic edges.
The reader can verify this claim by looking ahead to Equations~(\ref{eq:rcequiv})
and~(\ref{eq:Potts}).
Thus, if $G$ has a cut of size $b$
then
$$T(G;x',y') \geq
(y'-1)^n (x'-1)^{-\kappa(E)}
(y')^{m-b}.$$
Otherwise,
$$T(G;x',y') \leq (y'-1)^n (x'-1)^{-\kappa(E)}
2^n (y')^{m-b+1}.$$

Now consider a point $(x,y)$ on $H_2$ with $x<-1$. Note that $y\in(0,1)$.
Let $k$ be a
positive integer with $y^k<2^{-2n}$ and let $y'=y^k$.
Let $x'=2/(y^k-1)+1$.
If we had an FPRAS for $\tutte(x,y)$,
we could estimate $T(G;x',y')$ by $k$-thickening.
Thus, we could determine whether or not $G$ has a cut of
size~$b$, giving $\mathrm{RP}=\mathrm{NP}$.

\end{proof}

\begin{corollary}\label{cor:y<-1}
Suppose $(x,y)$ is a point lying in the open half-plane
$y<-1$ but not on the hyperbolas~$H_0$, $H_1$ or $H_{2}$.
Under the assumption $\mathrm{RP}\not=\mathrm{NP}$
there is no FPRAS for $\tutte(x,y)$.
\end{corollary}

\begin{proof}
Dual to the proof of Corollary~\ref{cor:x<-1}
(but without the extra argument for $q=2$).
\end{proof}

Corollaries~\ref{cor:x<-1} and~\ref{cor:y<-1}
exclude the hyperbola $q=0$.
Nevertheless, the arguments of Theorem~\ref{thm:shift}
can be extended to handle the portion of this
(degenerate) hyperbola in which $y<-1$.
Specifically, in Section~\ref{sec:reductions}
we  prove the following.

\begin{lemma}
\label{lem:x=1}
Suppose $(x,y)$ is a point with $x=1$ and $y<-1$.
Then there is no FPRAS for $\tutte(x,y)$ unless
$\mathrm{RP}=\mathrm{NP}$.
\end{lemma}

We do not know whether the arguments of Theorem~\ref{thm:shift'} can
be similarly extended to $q=0$.

The hyperbola $H_2$ is excluded from
Theorems~\ref{thm:shift} and \ref{thm:shift'} and a separate argument
(following Welsh) was used to include $H_2$ within the scope of Corollary~\ref{cor:x<-1}
which applies to the region $x<-1$. We do not know of a similar argument
which applies to $H_2$ in the region $y<-1$ and indeed
this hyperbola seems to have a special status in the region $y<-1$,
as Lemma~\ref{lem:matchings} shows.
Consider the following computational problem.
\begin{description}
\item[Name] \PM.
    \item[Instance]  A graph $G$.

    \item[Output] The number of perfect matchings in~$G$.
\end{description}
\PM\ is \#P-complete, but it is not know whether it has an FPRAS.
In Section~\ref{sec:reductions} we prove the following.

\begin{lemma}
\label{lem:matchings}
Suppose $(x,y)$ is a point on the hyperbola~$H_2$ with $y<-1$.
Then $\tutte(x,y) \APequiv \PM$.
\end{lemma}

\noindent{\bf Remark:\quad}
For convenience, we allow the graph $G$ in the definition of
\PM\ to have loops and/or multiple edges. This is without loss of
generality, since the perfect matchings of a graph $G$
are in one-to-one correspondence with the perfect matchings of
the $3$-stretch of $G$ (which has no loops or multiple edges).

\subsection{The Vicinity of the Origin}
\label{sec:vic}

In this section, we consider the region given by $|x|<1$ and $|y|<1$.
We have already seen (in the proof of Corollary~\ref{cor:x<-1})
that, unless $\mathrm{RP}=\mathrm{NP}$, there is no FPRAS for
$\tutte(x,y)$  for any point $(x,y)$ on the hyperbola
$(x-1)(y-1)=2$ in this region.
The following lemmas give additional regions that do not admit
an FPRAS unless $\mathrm{RP}=\mathrm{NP}$.

\begin{lemma}
\label{lem:triangle1}
Suppose $(x,y)$ is a point with $|x|<1$, $|y|<1$ and $y<-1-2x$
that does not
lie on the hyperbola~$H_{1}$.
Then
there is no FPRAS for $\tutte(x,y)$ unless $\mathrm{RP}=\mathrm{NP}$.
\end{lemma}
\begin{proof}
Using Equation~(\ref{eq:stretchThickenRules}),
a $2$-stretch shifts $(x,y)$ to $(x',y')$ with
$$y'=\frac{(x-1)(y-1)}{x^2-1}+1 = \frac{y+x}{x+1}<-1.$$
Now if $q=(x-1)(y-1)\notin\{0,1,2\}$, the lemma follows
from Theorem~\ref{thm:shift}.
As we noted above, the result is already known for $q=2$.
Also, $H_0$ is outside the scope
of the lemma.
\end{proof}

\begin{lemma}
\label{lem:triangle2}
Suppose $(x,y)$ is a point with $|x|<1$, $|y|<1$ and $x<-1-2y$
that does not satisfy $(x-1)(y-1)=1$. Then
there is no FPRAS for $\tutte(x,y)$ unless $\mathrm{RP}=\mathrm{NP}$.
\end{lemma}

\begin{proof}
As in the proof
of Lemma~\ref{lem:triangle1} we can use a $2$-thickening
together with Theorem~\ref{thm:shift'} to obtain the result when
except for $q=0$, $q=1$ and $q=2$. The result is known for $q=2$ and
the cases $q=0$ and $q=1$ are excluded from the  lemma.
\end{proof}

Lemmas~\ref{lem:triangle1} and~\ref{lem:triangle2}
give two intractable open triangles in the vicinity of
the origin.
The following lemmas
extend intractability to the boundaries.
The value $0.29$ in the statement of the lemmas has no
special significance. We do not know whether the
entire boundary is intractable, but the value $0.29$ is not
best possible --- it was chosen because it yields a simple proof.

\begin{lemma}
\label{lem:boundary1}
Suppose $(x,y)$ is a point with $x=-1$ and $-1<y<0.29$,
excluding the special point $(x,y)=(-1,0)$.
Then there is no FPRAS for $\tutte(x,y)$ unless
$\mathrm{RP}=\mathrm{NP}$.
\end{lemma}

\begin{proof}
A $2$-thickening of $(-1,y)$ gives the point $(x',y')=
((y-1)/(y+1),y^2)$. If $y<0$ then $x'<-1$ so the
result follows from Corollary~\ref{cor:x<-1} since
$(x',y')$ is not on $H_0$ or $H_1$.
Now if $0<y<1$ then $x'\in(-1,0)$ so $|x'|<1$ and $|y'|<1$.
Now note that if $0<y<0.29$ then
$y'<-1-2x'$ so the result follows from Lemma~\ref{lem:triangle1}.
\end{proof}

\begin{lemma}
\label{lem:boundary2}
Suppose $(x,y)$ is a point with $y=-1$ and $-1<x<0.29$,
excluding the special point $(x,y)=(0,-1)$.
Then there is no FPRAS for $\tutte(x,y)$ unless
$\mathrm{RP}=\mathrm{NP}$.
\end{lemma}

\begin{proof}
Dual to the proof of Lemma~\ref{lem:boundary1}.
\end{proof}

The intractable triangles from Lemma~\ref{lem:triangle1}
and~\ref{lem:triangle2} certainly do not cover
all intractable points in the vicinity of the origin.
Possibly the whole of the region $|x|,|y|\leq 1$
is intractable (apart from $H_1$ and the special points).

Here is a lemma which adds a little bit to our
knowledge in the region.
For example, it includes the point $(x,y)=(-0.23,-0.23)$ which
has $q>1.5$
but is not covered by Lemma~\ref{lem:triangle1}
or~\ref{lem:triangle2}.

\begin{lemma}
\label{lem:vicinity}
Suppose $(x,y)$ is a point with $|x|\leq 1$ and
$|y|\leq 1$ and $(x-1)(y-1)=q > 1.5$ (excluding the special points
$(-1,-1)$, $(-1,0)$ and $(0,-1)$).
Then there is no FPRAS for $\tutte(x,y)$ unless
$\mathrm{RP}=\mathrm{NP}$.
\end{lemma}

\begin{proof}

First, note that neither $x$ nor $y$ is $1$, since that would make $q=0$.
Also, we don't have $(x,y)=(-1,-1)$ since that is a special point.
Suppose $x=-1$. Then $y>-1$. The restriction on $q$ implies $y<0.25$,
so the result follows from Lemma~\ref{lem:boundary1}.
Similarly, the case $y=-1$ follows from Lemma~\ref{lem:boundary2}.
So suppose $|x|<1$ and $|y|<1$.

If $q>2$ then
the result follows from Theorem~\ref{thm:shift'}.
Do a $2$-thickening (Equation~(\ref{eq:stretchThickenRules}))
to shift to the point
$$(x',y')=\left(\frac{q}{y^2-1}+1,y^2\right).$$
Note that $y^2-1\in(-1,0)$ since $|y|<1$
so $q/(y^2-1)<-q<-2$  since $q>2$. So $x'<-1$.
Then  apply Theorem~\ref{thm:shift'}.
The case $q=2$ is known, as noted at the beginning of the section.
Suppose $3/2<q<2$.
For a large even integer~$k$,
do a $k$-thickening to shift $(x,y)$ to the point
$$(x_1,y_1) =\left(\frac{q}{y^k-1}+1,y^k\right).$$
Choose $k$ so that $0<y^k <(2-q)/2$ (this is possible since $q<2$).
Consider $x_1 - 1 = q/(y^k-1)$. Note that this is in the interval $(-2,-q)$.
Now do a $2$-stretch to shift $(x_1,y_1)$ to
the point
$(x',y')=(x_1^2,q/(x_1^2-1)+1)$.
Note that
$$y'-1 = \frac{q}{x_{1}^2-1}<-2$$
where the upper bound of $-2$ follows from the bounds that
we derived on $x_{1}$ and $q>3/2$.
Now use Theorem~\ref{thm:shift}.
\end{proof}

The lemma could certainly be improved.
For example, consider the point $(x,y) = (-0.2,0)$
with $q = 1.2$. An alternating sequence of
$14$ $2$-stretches and $2$-thickenings
shifts this point to a point $(x',y')\sim(-103.1,0.99)$
so $(x,y)$ had no FPRAS (unless $\mathrm{RP}=\mathrm{NP}$)
by Theorem~\ref{thm:shift'}.

\section{The reductions}\label{sec:reductions}

\subsection{The Multivariate Formulation of the Tutte Polynomial}

It is convenient
for us to use the multivariate formulation of the
Tutte polynomial, also known as the random cluster model~\cite{Welsh93, Sokal05}.
For a graph $G=(V,E)$ with edge weights
$w:E\to\mathbb{Q}$
and $q\in\mathbb{Q}$,
define the multivariate Tutte polynomial of $G$ to be
$Z(G;q,w)=\sum_{A\subseteq E}w(A)q^{\kappa(A)}$,
where $w(A)=\prod_{e\in A}w(e)$, and
$\kappa(A)$ is the number of connected components in the
graph $(V,A)$.

Suppose $(x,y)\in\mathbb{Q}^2$ and $q=(x-1)(y-1)$.
For a graph $G=(V,E)$, let $w:E\to\mathbb{Q}$ be the constant
function which maps every edge to the value $y-1$.
Then (see, for example \cite[(2.26)]{Sokal05})
\begin{equation}
T(G;x,y) = {(y-1)}^{-n}{(x-1)}^{-\kappa(E)} Z(G;q,w).
\label{eq:rcequiv}
\end{equation}

So $Z$ is a generalisation of $T$ that allows different weights
to be assigned independently to different edges.
For rationals $q$ and $\gamma$,
let $\multitutte(q,\gamma)$ be the
following problem.
\begin{description}
\item [Name] $\multitutte(q,\gamma)$.
    \item[Instance]  A graph $G=(V,E)$ with
    edge labelling $w$ where $w$ is the constant function
    mapping every edge to the value $\gamma$.
    \item[Output]  $Z(G;q,w)$.
\end{description}

Suppose $(q,\gamma)\in {\mathbb{Q}}^2$. Equation~(\ref{eq:rcequiv})
gives us the reduction
$$\multitutte(q,\gamma)\APred \tutte\left(\tfrac{q}{\gamma}+1,\gamma+1\right).$$
If $(x,y)\in {\mathbb{Q}}^2$ and neither~$x$ nor~$y$ is~$1$
then Equation~(\ref{eq:rcequiv}) gives us the corresponding
reduction
$$\tutte(x,y)\APred \multitutte((x-1)(y-1),y-1).$$

Not surprisingly, the notion of a shift from~\S\ref{sec:map}
may be re-expressed in terms of the new parameters.
Doing so has the advantage of allowing us to apply shifts
to individual edges of a graph, as opposed to the whole
graph.
This idea is explored in~\cite[\S4.6]{Sokal05}. We derive the
equations that we need here in order to fix the notation and
explore the concepts.
As in \S\ref{sec:map}, let $K$ be a graph with distinguished
edge~$e$, and suppose that $K$ has constant edge
weight~$\alpha\in\mathbb{Q}$.  Define
\begin{align}
\alpha'&=\frac{q\,Z(K/e;q,\alpha)-Z(K\setminus e;q,\alpha)}
   {Z(K\setminus e;q,\alpha)-Z(K/e;q,\alpha)}\label{eq:alpha'defn}\\
\noalign{\noindent and}
N(q,\alpha,K)&=\frac{q(q-1)}
{Z(K\setminus e;q,\alpha)-Z(K/e;q,\alpha)}.\notag
\end{align}
Let $G=(V,E)$ be a graph, $f\in E$ and $w':E\to\mathbb{Q}$ a weighting
such that $w'(f)=\alpha'$.  Denote by $G_{f}$ the 2-sum of $G$ and~$K$
along~$f$.
Let $w$ be the weight function on~$G_f$ that gives every edge of~$K$
weight~$\alpha$ and inherits the remaining weights from~$w'$.
We will show below that
\begin{equation}\label{eq:2sum}
    Z(G;q,w')= N(q,\alpha,K)\, Z(G_{f};q,w).
\end{equation}
One way to capture~(\ref{eq:2sum}) informally is to say that
a single edge of weight~$\alpha'$ may be simulated by a
subgraph~$K$  whose edges have weight~$\alpha$.

Suppose that the denominator of (\ref{eq:alpha'defn}) is non-zero.
In this case,
the point $(q,\alpha')$ is well-defined and we say that
$(q,\alpha)$ is {\it shifted to\/} the point $(q,\alpha')$ by~$K$.
In this case, $N(q,\alpha,K)$ is also well-defined,
so Equation~(\ref{eq:2sum}) gives us
an efficient algorithm for approximating $Z(G;q,w')$ by using
an subroutine for computing $Z(G_f;q,w)$.

For the derivation of (\ref{eq:alpha'defn})
and (\ref{eq:2sum}), let $v$ and $v'$ be the endpoints of~$f$.
Let $S$ be the set of subsets of $E-\{f\}$ which connect $v$ and~$v'$ and
let $T$ be the set of all other subsets of $E-\{f\}$.
Then $Z(G;q,w')= Z_S+Z_T$,
where
$$Z_S = \sum_{A'\in S} w(A')q^{k(A')}(1+\alpha'),$$
and
$$Z_T = \sum_{A'\in T} w(A')q^{k(A')}\left(1+\tfrac{\alpha'}{q}\right).$$
Similarly,
$Z(G_f,q,w) = Z_{f,S}+Z_{f,T}$,
where
$$Z_{f,S} =\sum_{A'\in S} w(A')q^{k(A')}\frac{Z(K/e;q,\alpha)}{q},$$
and
$$Z_{f,T} = \sum_{A'\in T} w(A')q^{k(A')}\frac{Z(K\setminus e;q,\alpha)}{q^2}.$$
Now the equation for~$\alpha'$ comes
from the
following argument. Suppose we could define $\alpha'$ such that
\begin{equation}
\label{eqlast}
\frac{Z_S}{Z_T} = \frac{Z_{f,S}}{Z_{f,T}},
\end{equation}
and $N(q,\alpha,K) = Z_S/Z_{f,S}$.
Then (\ref{eq:2sum}) would hold as desired. Now note that (\ref{eq:alpha'defn}) entails
(\ref{eqlast}).

The shifts that we have defined here
are consistent with the usage in \S\ref{sec:map}.
In particular, suppose that $(x,y)$ is shifted to the point $(x',y')$
by a graph~$K$ with distinguished edge~$e$. As long as $e$ is not
a bridge of~$K$ then taking $\alpha=y-1$ and $\alpha'=y'-1$ and
$q=(x-1)(y-1)$ we find (from Equations~(\ref{eq:y'defn})
and (\ref{eq:rcequiv}) and (\ref{eq:alpha'defn})) that the same graph~$K$
shifts $(q,\alpha)$ to $(q,\alpha')$.

Thus, the equation describing stretching and thickening,
Equation~(\ref{eq:stretchThickenRules}), can be translated as follows.
(See, for example,   \cite[(4.20), (4.26)]{Sokal05})
\begin{equation}
\begin{aligned}
\frac{q}{\alpha'}&=\left(\frac{q}{\alpha}+1\right)^{k}-1,
   &\text{for a $k$-stretch;}\\
\alpha'&=(\alpha+1)^{k}-1,&\text{for a $k$-thickening.}
\label{eq:stretchThickenRules'}
\end{aligned}
\end{equation}

We now generalise the computational problem $\multitutte(q,\gamma)$ defined
earlier.
For rationals $q,\alpha_{1},\ldots,\alpha_{k}$,
$\multitutte(q;\alpha_{1},\ldots,\alpha_{k})$ is the
problem:
\begin{description}

\item [Name] $\multitutte(q;\alpha_1,\ldots,\alpha_k)$.
    \item[Instance]  A graph $G=(V,E)$ with
    edge labelling $w:E\to\{\alpha_{1},\ldots,\alpha_{k}\}$.

    \item[Output]  $Z(G;q,w)$.
\end{description}

\subsection{Proof of Theorem~\ref{thm:shift}}

The decision problem \textsc{Minimum 3-way Cut} is:
\begin{description}
    \item[Instance]  A simple graph $G=(V,E)$ with
    three distinguished vertices (``terminals'') $t_1,t_2,t_3\in V$,
    and an integer bound~$b$.

    \item[Output]  Is there a set of at most $b$ edges whose
    removal from~$G$ disconnects $t_i$ from~$t_j$ for every
    $i,j\in\{1,2,3\}$ with $i\not=j$?
\end{description}
It was shown to be NP-complete by Dahlhaus et al.~\cite{DJPSY94}.

\begin{lemma}\label{lem:bedrock}
    Suppose $q\in \mathbb{Q}\setminus\{0,1,2\}$, and that
    $\alpha_1,\alpha_2\in\mathbb{Q}$ satisfy $\alpha_1\notin[-2,0]$
    and $\alpha_2\in(-2,0)$.  Then there is no FPRAS for
    $\multitutte(q;\alpha_{1},\alpha_{2})$ unless $\mathrm{RP}=\mathrm{NP}$.
\end{lemma}

\begin{proof}
Suppose $G=(V,E,t_{1},t_{2},t_{3})$ is an arbitrary instance
of \textsc{Min 3-way Cut}.  Without loss of generality assume
$G$ is connected, and
for convenience let $n=|V|$ and $m=|E|$.
Our ultimate goal is to construct an instance $(G',w')$ of
$\multitutte(q;\allowbreak\alpha_{1},\alpha_{2})$ such that $Z(G';q,w')$
is a close approximation to the number of minimum 3-way cuts
in~$G$.  The size of a minimum cut will be a by-product of the
of the reduction.

As an intermediate goal, we'll construct a weighted graph
$(\Ghat=(\Vhat,\Ehat),w)$
such that $Z(\Ghat;q,w)$ is a close approximation to the number of
minimum 3-way cuts in~$G$ where
$w:\Ehat\to\{\beta_{1},\beta_{2}\}$ for some conveniently-chosen
values
$\beta_{1}$ and $\beta_{2}$.  The final step of the proof will be
to relate these convenient values to the specified ones,
namely $\alpha_{1}$ and~$\alpha_{2}$.
We will require $\beta_1$ to be sufficiently large,
in particular,
let $\overline{q}=\max(|q|,1)$.
Let $M=8\times 2^m {\overline{q}}^n$.
We will require
\begin{equation}
\beta_1\geq M.
\label{eq:Mbd}
\end{equation}
We will also require $\beta_2$ to be sufficiently close to~$-1$.
In particular,
we will choose a small value~$\delta$ (see Equations~(\ref{eq:deltaBd1})
and~(\ref{eq:deltaBd2}), depending on $m$, $q$ and $n$.
We will require $|1+\beta_2|\leq \delta$.

The construction of $(\Ghat=(\Vhat,\Ehat),w)$ is very direct:
$\Vhat=V$,
$\Ehat=E\cup T$, where
$T=\big\{\{t_{1},t_{2}\},\{t_{2},t_{3}\},\{t_{1},t_{3}\}\big\}$,
and
$$
w(e)=\begin{cases}
    \beta_{1},&\text{if $e\in E$};\\
    \beta_{2},&\text{otherwise}.
\end{cases}
$$
Now, letting
$$
\calA_{1|2,3}=\{A\subseteq E: t_{1}\not\sim_{A} t_{2}\text{ and }
    t_{1}\not\sim_{A} t_{3}\text{ and }t_{2}\sim_{A} t_{3}\big\},
$$
etc, where $\sim_A$ denotes the binary relation ``is connected to''
in the graph $(V,A)$,
we may express the multivariate Tutte polynomial of~$\Ghat$ as
\begin{equation}\label{eq:sumPartition}
Z(\Ghat;q,w)=\subsum_{1|2|3}+\subsum_{1|2,3}
   +\subsum_{2|1,3}+\subsum_{3|1,2}+\subsum_{1,2,3},
\end{equation}
where, e.g.,
$$
\subsum_{1|2,3}=
   \sum_{A\in\calA_{1|2,3}}\sum_{B\subseteq T}w(A\cup B)q^{\kappa(A\cup B)}.
$$

The overview of the proof is as follows: we show that for
$\beta_{1}$  sufficiently large
and $\beta_{2}$ sufficiently close to~$-1$, the last four terms
on the r.h.s.\ of~(\ref{eq:sumPartition}) are negligible in comparison
with the first, and that the first term, $\subsum_{1|2|3}$,
counts minimum 3-way cuts in~$G$ (approximately, and up to
an easily computable factor).
Up to symmetry there are three essentially distinct terms
in~(\ref{eq:sumPartition}),
and we consider them in turn.  First,
\begin{align}
|\subsum_{1,2,3}|&=\Biggl|\sum_{A\in\calA_{1,2,3}}
   \sum_{B\subseteq T}w(A\cup B)q^{\kappa(A\cup B)}\Biggr|\notag\\
&=\Biggl|\sum_{A\in\calA_{1,2,3}}
   w(A)q^{\kappa(A)}(1+\beta_{2})^{3}\Biggr|\notag\\
&\leq (2\beta_1)^{m}|q|\,\qbar^{\,n-1}\delta^{3},\label{eq:sumEstimate1}
\end{align}
Here we have used $1\leq\kappa(A)\leq n$.
Next,
\begin{align}
|\subsum_{1|2,3}|&=\Biggl|\sum_{A\in\calA_{1|2,3}}
   \sum_{B\subseteq T}w(A\cup B)q^{\kappa(A\cup B)}\Biggr|\notag\\
&=\Biggl|\sum_{A\in\calA_{1|2,3}}
   w(A)q^{\kappa(A)-1}
   (q+2\beta_{2}+\beta_{2}^{2})(1+\beta_{2})\Biggr|\notag\\
&\leq (2\beta_1)^{m}|q|\,\qbar^{\,n-2}\,
(\overline{q}+4+4)\delta\notag\\
&\leq 9(2\beta_1)^{m}|q|\,\qbar^{\,n-1}\delta.\label{eq:sumEstimate2}
\end{align}
Here we used $|\beta_{2}|\leq2$ and $2\leq\kappa(A)\leq n$.  Last,
\begin{align}
\subsum_{1|2|3}&=\sum_{A\in\calA_{1|2|3}}
   \sum_{B\subseteq T}w(A\cup B)q^{\kappa(A\cup B)}\notag\\
&=\sum_{A\in\calA_{1|2|3}}w(A)q^{\kappa(A)-2}
   \big(q^{2}+3\beta_{2}q+3\beta_{2}^2+\beta_{2}^3\big)\notag\\
&=C(\beta_{2})\sum_{A\in\calA_{1|2|3}}w(A)q^{\kappa(A)-2},\label{eq:sumEstimate3}
\end{align}
where
$$
C(\beta_{2})=(q-1)(q-2)+3(q-1)(1+\beta_{2})+(1+\beta_{2})^{3}.
$$
Note that
$$
\bigl|C(\beta_{2})-(q-1)(q-2)\bigr|\leq3\,|q-1|\,\delta+\delta^{3},
$$
The crucial fact is that $C(\beta_{2})$ remains bounded
away from 0 as $\delta\to0$ (and hence $\beta_2\to-1$),
provided (as we are assuming) $q\notin\{1,2\}$, whereas
expressions (\ref{eq:sumEstimate1}) and~(\ref{eq:sumEstimate2})
tend to~0 as $\delta\to0$.

Now denote by $\calA_{1|2|3}^{(i)}$ the set of all subsets in $\calA_{1|2|3}$
of size~$i$.  Let $c$ be the size of a minimum $3$-way cut in~$G$, and $N$ be
the number of such cuts. Then

\begin{align*}
 \frac{\subsum_{1|2|3}}{C(\beta_{2})}
   &=\sum_{A\in\calA_{1|2|3}^{(m-c)}}w(A)q^{\kappa(A)-2}+
   \sum_{i=0}^{m-c-1}\sum_{A\in\calA_{1|2|3}^{(i)}}w(A)q^{\kappa(A)-2}\\
&=N\beta_1^{m-c}q+
   \sum_{i=0}^{m-c-1}\sum_{A\in\calA_{1|2|3}^{(i)}}w(A)q^{\kappa(A)-2}.
\end{align*}

Then
$$\frac{\subsum_{1|2|3}}{C(\beta_{2})\beta_1^{m-c}q}-N$$
is equal to
$$\Bigg(
\sum_{i=0}^{m-c-1}\sum_{A\in\calA_{1|2|3}^{(i)}}w(A)q^{\kappa(A)-2}
\Bigg)\Bigg/\left(\beta_1^{m-c}q\right),$$
so crudely upper-bounding the absolute value of the right-hand-side,
we get
\begin{equation}\label{eq:Napprox}
\left|\frac{\subsum_{1|2|3}}{C(\beta_{2})\beta_1^{m-c}q}-N\right|
\leq \frac{2^{m}\qbar^{\,n}}{\beta_1}.
\end{equation}
Now set $\delta$ to satisfy both
\begin{equation}
|C(\beta_{2})|\geq\tfrac12|C(-1)|\label{eq:deltaBd1}
\end{equation}
and
\begin{equation}
\delta\leq\frac{|C(-1)|}{448(2\beta_1)^{m}\qbar^{\,n}}.\label{eq:deltaBd2}
\end{equation}
Now (\ref{eq:Mbd}) ensures that
the r.h.s. of~(\ref{eq:Napprox}) is at most $\frac18$,
while (\ref{eq:sumEstimate1}), (\ref{eq:sumEstimate2}), (\ref{eq:sumEstimate3}),
(\ref{eq:deltaBd1}) and (\ref{eq:deltaBd2}) ensure
$$
\left|\frac{\subsum_{1|2|3}-Z(\Ghat;q,w)}
   {C(\beta_{2})\beta_1^{m-c}q}\right|=
\left|\frac{\subsum_{1|2,3}+\subsum_{2|1,3}+\subsum_{3|1,2}+\subsum_{1,2,3}}
   {C(\beta_{2})\beta_1^{m-c}q}\right|\leq\frac18.
$$
Combining this inequality with (\ref{eq:Napprox}), the bottom line is

\begin{equation}\label{eq:Napprox'}
\left|\frac{Z(\Ghat;q,w)}{C(\beta_{2})\beta_1^{m-c}q}-N\right|
\leq \frac14.
\end{equation}

If we knew $Z(\Ghat;q,w)$, we could determine $c$ --- it
is the unique integer such that
(\ref{eq:Napprox'}) provides an estimate for $N$ that
lies in the range $[1,2^{m}]$. The value of $c$ is unique since
$\beta_1 \geq M >2^m$.)

In fact, we do not need an exact value of $Z(\Ghat;q,w)$ ---
an approximate value will do.
In particular, an FPRAS for $Z(\Ghat;q,w)$
would give a randomised polynomial-time algorithm
for computing $c$, which would show $\mathrm{RP}=\mathrm{NP}$.
For details about approximation accuracy, see
\cite{DGGJ04}, especially the final three paragraphs of
the proof of Theorem~3.

Finally we need to relate our weights $\beta_{1},\beta_{2}$
to the given ones $\alpha_{1},\alpha_{2}$.  Let positive
integers $k_{1},k_{2}$ satisfy $(\alpha_{1}+1)^{k_{1}}-1\geq M$
and $\bigl|(\alpha_{2}+1)^{k_{2}}\bigr|<\delta$.
Let $K_1$ be a $2$-vertex graph with $k_1+1$ parallel edges, each
of weight $\alpha_1$. Recall that taking a $2$-sum with
$K_1$ implements a $k_1$ thickening.
Let $K_2$ be
a $2$-vertex graph with $k_2+1$ parallel edges, each of weight $\alpha_2$.
Let $G'$ be the graph derived from~$\Ghat$ by
taking the $2$-sum of each weight $\beta_1$ edge with $K_1$
and taking the $2$-sum of each weight $\beta_2$ edge with $K_2$.
Call the resulting graph $G'$ and its weighting~$w'$.
By repeated application of~(\ref{eq:2sum}),
$Z(\Ghat;q,w)={N(q,\alpha_1,K_1)}^m {N(q,\alpha_2,K_2)}^3Z(G';q,w')$.
By setting $\beta_{i}=(\alpha_{i}+1)^{k_{i}}-1$, for $i=1,2$,
we satisfy $\beta_{1}>M$ and $|\beta_{2}+1|\leq\delta$,
as required by our reduction.
(This is by~(\ref{eq:stretchThickenRules'}) and the definitions
of $k_{1}$ and~$k_{2}$.)  Finally observe that $k_{1}=O(m)$
and $k_{2}=O(m^{2})$, so the size of~$G'$ is polynomially bounded.

Thus an FPRAS for $\multitutte(q;\alpha_{1},\alpha_{2})$ would yield
a polynomial-time randomised algorithm for computing the size of
a minimum 3-way cut, which would entail
$\mathrm{RP}=\mathrm{NP}$.
\end{proof}

Using Lemma~\ref{lem:bedrock}, we can now prove
Theorem~\ref{thm:shift}.

\noindent{\bf Theorem \ref{thm:shift}}
{\sl
Suppose $(x,y)\in \mathbb{Q}^2$ satisfies
$q=(x-1)(y-1)\notin\{0,1,2\}$.  Suppose also that
it is possible to shift the point
$(x,y)$ to the point $(x',y')$ with $y'\notin[-1,1]$,
and to $(x'',y'')$ with $y''\in(-1,1)$.
Then there is no FPRAS for
$\tutte(x,y)$
unless
$\mathrm{RP}=\mathrm{NP}$.
}

\begin{proof}
Let $\alpha=y-1$ and $\alpha_1=y'-1$ and $\alpha_2=y''-1$.
Note that $\alpha_1\notin[-2,0]$ and $\alpha_2\in(-2,0)$.
Let $(K',e')$ be a graph that shifts $(x,y)$ to
$(x',y')$ and note that $(K',e')$ also shifts $(q,\alpha)$ to $(q,\alpha_1)$.
Similarly, suppose $(K'',e'')$ shifts $(x,y)$ to $(x'',y'')$ and
therefore shifts $(q,\alpha)$ to $(q,\alpha_2)$.

Suppose $(G,w)$ is an instance of $\multitutte(q;\alpha_{1},\alpha_{2})$
with $m_1$ edges with weight $\alpha_1$ and $m_2$ edges with weight $\alpha_2$.
Denote by $\hatG$ the graph derived from $G$ by taking a 2-sum with
$(K',e')$ along every edge with weight $\alpha_1$ and taking a 2-sum with
$(K'',e'')$ along every edge with weight $\alpha_2$.
Let $\hat{w}$ be the constant weight function which assigns weight $\alpha$ to
every edge in $\hatG$.

Then by repeated use of Equation~(\ref{eq:2sum}),
$$Z(G;q,w) = {N(q,\alpha,K_1)}^{m_1} {N(q,\alpha,K_2)}^{m_2}
Z(\hatG;q,\hat{w}).$$

Thus by Equation~(\ref{eq:rcequiv}),
$$Z(G;q,w) = {N(q,\alpha,K_1)}^{m_1} {N(q,\alpha,K_2)}^{m_2}
(y-1)^n(x-1)^\kappa
T(\hatG;x,y),$$
where $n$ is the number of vertices in $\hatG$, and $\kappa$ is the number of
connected components in $\hatG$.

Thus an FPRAS for $\tutte(x,y)$
would yield an FPRAS
for the problem
$\multitutte(q;\alpha_{1},\alpha_{2})$, contrary to Lemma~\ref{lem:bedrock}.
\end{proof}

\subsection{Extending to $q=0$}

Formally, the multivariate Tutte polynomial
$Z(G;q,w)=\sum_{A \subseteq E}w(A)q^{\kappa(A)}$ is not
very interesting at $q=0$ because $\kappa(A)\geq 1$ so $Z(G;q,w)=0$.
Sokal~\cite{Sokal05} treats the $q=0$ case as a limit, but
for the purpose of approximation complexity it is more convenient to
work with the polynomial
$Z(G;q,w)q^{-\kappa(E)}$. We will focus on the case in which~$G$ is connected,
so we define
$R(G;q,w)=Z(G;q,w)q^{-1}$. Note that
\begin{equation}
\label{eq:Rdefn}
R(G;0,w)=\sum_{A\subseteq E: \kappa(A)=1}w(A).
\end{equation}
This is the reliability polynomial, and corresponds to the $x=1$
component of the hyperbola~$H_{0}$.

We can express shifts in terms of $R(G;q,w)$.
For example, Equation~(\ref{eq:stretchThickenRules'}) does not tell us anything useful
about stretching for $q=0$ (due to cancellation) but the same
reasoning that we used to derive~(\ref{eq:alpha'defn}) and (\ref{eq:2sum})
gives us the following version
of these equations for the case in which $G_f$ is a $k$-stretch (specifically,
$G_f$ is the $2$-sum of $G$ and a cycle on $k+1$ vertices along edge~$f$):
\begin{equation}
\label{newstretchalpha'}
\alpha'=\frac{\alpha}{k}
\end{equation}
and
\begin{equation}
\label{newstretch2sum}
R(G;0,w')= \frac{1}{k \alpha^{k-1}} R(G_f;0,w).
\end{equation}
As in the general case, we assume $w'$ is a weight function on $G$
with $w(f)=\alpha'$ and that $w$ inherits its weights from $w'$
except that the new edges in the stretch are given weight~$\alpha$.
The derivation of (\ref{newstretchalpha'}) and (\ref{newstretch2sum}) follows
the derivation of~(\ref{eq:alpha'defn}) and (\ref{eq:2sum}).
Specifically, let $S$ (respectively, $T$)
be the set of all subsets~$A'\subseteq E-\{f\}$
with $\kappa(A')=1$ (respectively, $\kappa(A')=2$ and $\kappa(A'\cup\{f\})=1)$.
Then
$R(G;0,w')=R_S + R_T$
where
\begin{align}
R_S&= \sum_{A'\subseteq S}w(A')(1+\alpha'),\\
R_T&= \sum_{A'\subseteq T}w(A')\alpha';
\end{align}
and $R(G_f;0,w')=R_{f,S} + R_{f,T}$, where
\begin{align}
R_{f,S}&= \sum_{A'\subseteq S}w(A')(\alpha^k+k\alpha^{k-1}),\\
R_{f,T}&= \sum_{A'\subseteq T}w(A')\alpha^k.
\end{align}

Similarly, for the case in which $G_f$ is a $k$-thickening, we
get
\begin{equation}
\label{newthicken2sum}
R(G;0,w')=R(G_f;0,w),
\end{equation}
with $\alpha'$ as in Equation~(\ref{eq:stretchThickenRules'})

Now let $\zerotutte(\alpha_1,\ldots,\alpha_k)$ be the
following problem.
\begin{description}

\item [Name] $\zerotutte(\alpha_1,\ldots,\alpha_k)$.
    \item[Instance]  A connected graph $G=(V,E)$ with
    edge labelling $w:E\to\{\alpha_{1},\ldots,\alpha_{k}\}$.

    \item[Output]  $R(G;0,w)$.
\end{description}

An examination of the proof of Lemma~\ref{lem:bedrock}
gives the following lemma.
\begin{lemma}\label{lem:zerobedrock}
    Suppose that
    $\alpha_1,\alpha_2\in\mathbb{Q}$ satisfy $\alpha_1\notin[-2,0]$
    and $\alpha_2\in(-2,0)$.  Then there is no FPRAS for
    $\zerotutte(\alpha_{1},\alpha_{2})$ unless $\mathrm{RP}=\mathrm{NP}$.
\end{lemma}

The proof of Lemma~\ref{lem:zerobedrock} follows that of Lemma~\ref{lem:bedrock}.
By analogy to Equation~(\ref{eq:sumPartition})
we may express $R(\hatG;0,w)$ as a sum of terms of the form
$\subsum_{1|2|3}$.
Then
$$
\subsum_{1|2|3}
= \sum_{A\in \mathcal{A}_{1|2|3}:\kappa(A)=3} w(A) (\beta_2^3  +
3 \beta_2^2),$$
and the other terms all have factors of~$\delta$.
By analogy to Equation~(\ref{eq:Napprox}) we
get
\begin{equation}
\left|\frac{\subsum_{1|2|3}}{(\beta_2^3+3\beta_2^2)\beta_1^{m-c}}-N\right|
\leq \frac{2^{m}}{\beta_1}.
\end{equation}

Using Lemma~\ref{lem:zerobedrock},
we can now prove Lemma~\ref{lem:x=1}.

\noindent{\bf Lemma
\ref{lem:x=1}.\quad}{\sl
Suppose $(x,y)$ is a point with $x=1$ and $y<-1$.
Then there is no FPRAS for $\tutte(x,y)$ unless
$\mathrm{RP}=\mathrm{NP}$.
}

\begin{proof}
Let $(x,y)$ be a point with $x=1$ and $y<-1$.
Let $\alpha=y-1$ and $q=0$.
Note that $\alpha \notin[-2,0]$.
Let $k=\lfloor -\alpha \rfloor$ and let $\alpha_2=\alpha/k$.
Note that $\alpha_2\in(-2,0)$,
and by Equation~(\ref{newstretch2sum}),
a $k$-stretch shifts $(q=0,\alpha)$ to
$(q=0,\alpha_2)$.
Suppose $(G,w)$ is an instance of
$\zerotutte(\alpha,\alpha_2)$ with $m_2$
edges with weight $\alpha_2$.
Denote by $\hatG$ the graph derived from $G$
by applying a $k$-stretch to each of these $m_2$~edges.
Let $\hat{w}$ be the constant weight function
which assigns weight $\alpha$ to every edge in $\hatG$.
Then by repeated use of Equation~(\ref{newstretch2sum}),
$$R(G;0,w) = {\left(\frac{1}
{k \alpha^{k-1}}\right)}^{m_2} R(\hatG;0,\hat{w}).$$
Using Equation~(\ref{eq:Rdefn}),
$$R(G;0,w) = {\left(\frac{1}
{k \alpha^{k-1}}\right)}^{m_2}
\sum_{A \subseteq E:\kappa(A)=1} {(y-1)}^{|A|},$$
where $E$ is the edge set of $\hatG$, which is connected
since $G$ is.
Thus, by the definition of the Tutte polynomial
(\ref{eq:tutte}),
$$R(G;0,w)= {\left(\frac{1}
{k \alpha^{k-1}}\right)}^{m_2}
{(y-1)}^{n-1}
T(\hatG;x,y),$$
where $n$ is the number of vertices of $\hatG$.
So an FPRAS for $\tutte(x,y)$ would enable us
to approximate $R(G;0,w)$, contrary to
Lemma~\ref{lem:zerobedrock}.

\end{proof}

\subsection{Proof of Theorem~\ref{thm:shift'}}

The following is dual to Lemma~\ref{lem:bedrock}.

\begin{lemma}\label{lem:dual}
    Suppose $q\in \mathbb{Q}\setminus\{0,1,2\}$, and that
    $\alpha_1,\alpha_2\in\mathbb{Q}-\{0\}$ satisfy
$q/\alpha_1 \notin[-2,0]$ and $q/\alpha_2 \in (-2,0)$.
Then there is no FPRAS for
    $\multitutte(q;\alpha_{1},\alpha_{2})$ unless $\mathrm{RP}=\mathrm{NP}$.
\end{lemma}
\begin{proof}

We reuse the construction that Frederickson and Ja' Ja'
designed in order to prove that
\textsc{Connected Bridge-connectivity Augmentation} (CBRA)
is NP-complete~\cite[Thm 2]{FredJaJa81},
though we'll change the edge weights to suit our purpose.
For convenience the construction will be repeated here.
We start with an instance of the 3-d Matching Problem:
$W$, $X$ and $Y$ are disjoint $n$-element sets, and
$M\subseteq W\times X\times Y$ a set of triples.
We want to know how many ``3-d matchings'' there are in~$M$.
A {\it 3-d matching\/} is a subset $M'\subseteq M$ of $n$~triples
such that every element of $W\cup X\cup Y$ is included in
some triple in~$M'$.  For convenience, we'll enumerate the
elements of the ground set $W=\{w_{1},\ldots,w_{n}\}$,
$X=\{x_{1},\ldots,x_{n}\}$,
and $Y=\{y_{1},\ldots,y_{n}\}$.\footnote{We'll stick, as
far as possible, to the notation of~\cite{FredJaJa81},
though occasional changes are needed to avoid clashes.}

Our ultimate goal is to construct an instance $(G',w')$ of
$\multitutte(q;\alpha_{1},\alpha_{2})$ such that $Z(G';q,w')$
is determined, to a high degree of accuracy, by the number of solutions
to the instance of \#\textsc{3-d Matching}.
In particular,
using an estimate of $Z(G';q,w')$,
we'll be able to decide, with high probability,
whether the number of solutions to the matching instance
is zero or strictly positive.
As an intermediate goal, just as in the proof of Lemma~\ref{lem:bedrock},
we'll construct a weighted graph
$(G=(V,E),w)$ that has the desired properties,
as described above,
except that $w:V\to\{\beta_{1},\beta_{2}\}$,
where $\beta_{1}$ and $\beta_{2}$ are set to
convenient non-zero values.  The final step of the proof will be
to relate these convenient values to the specified ones,
namely $\alpha_{1}$ and~$\alpha_{2}$.
The requirements on $\beta_1$ and $\beta_2$ are similar to the ones that
we used in the proof of Lemma~\ref{lem:bedrock}.
In particular, we will require, for a small $\epsilon\leq 1$,
that $|\beta_1/q| \leq \epsilon$ (so the absolute value of $q/\beta_1$
is big).
We will also require for a small $\delta\leq \tfrac12$ that
$|1+q/\beta_2|\leq \delta$ (so $\beta_2$ is close to $-q$).
We will require $\epsilon$ and $\delta$ to be sufficiently small ---
the exact requirements will be given later.

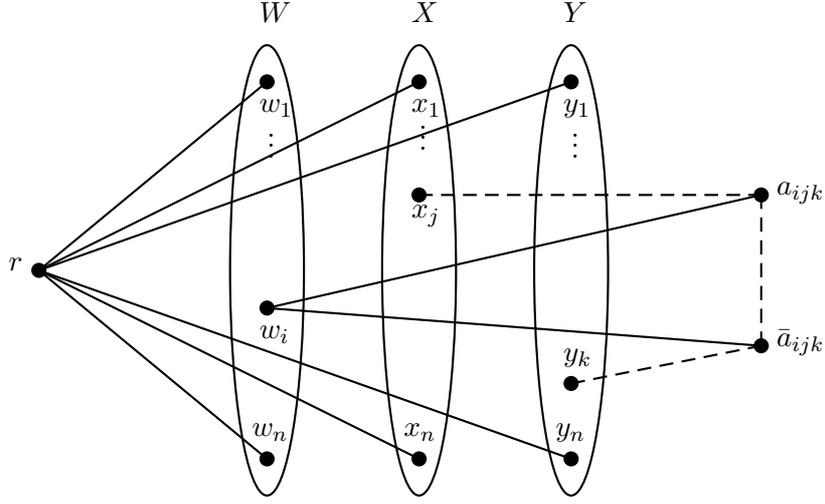
\begin{figure}[h]
\centering
\psset{unit=01}

\begin{pspicture}(0,2)(12,8.5)

\psellipse(6,5)(0.5,3)
\psellipse(4,5)(0.5,3)
\psellipse(8,5)(0.5,3)

\put(3.9,8.3){$W$}
\put(5.9,8.3){$X$}
\put(7.9,8.3){$Y$}

\pscircle*(1,5){0.1}
\put(0.6,5){$r$}

\pscircle*(4,7.5){0.1}
\pscircle*(4,2.5){0.1}
\put(3.9,7.1){$w_1$}
\put(3.8,2.8){$w_n$}
\put(4,6.5){$\vdots$}
\qline(1,5)(4,7.5)
\qline(1,5)(4,2.5)
\pscircle*(4,4.5){0.1}
\put(3.9,4.1){$w_i$}
\qline(4,4.5)(10.5,4)
\qline(4,4.5)(10.5,6)

\pscircle*(6,7.5){0.1}
\pscircle*(6,2.5){0.1}
\put(5.9,7.1){$x_1$}
\put(5.8,2.8){$x_n$}
\put(6,6.6){$\vdots$}
\qline(1,5)(6,7.5)
\qline(1,5)(6,2.5)
\pscircle*(6,6){0.1}
\put(5.9,5.7){$x_j$}
\psline[linestyle=dashed]{-}(6,6)(10.5,6)

\pscircle*(8,7.5){0.1}
\pscircle*(8,2.5){0.1}
\put(7.9,7.1){$y_1$}
\put(7.8,2.8){$y_n$}
\put(8,6.5){$\vdots$}
\qline(1,5)(8,7.5)
\qline(1,5)(8,2.5)
\put(7.9,3.8){$y_k$}
\pscircle*(8,3.5){0.1}
\psline[linestyle=dashed]{-}(8,3.5)(10.5,4)

\pscircle*(10.5,4){0.1}
\pscircle*(10.5,6){0.1}
\psline[linestyle=dashed]{-}(10.5,4)(10.5,6)
\put(10.7,6){$a_{ijk}$}
\put(10.7,4){$\bar a_{ijk}$}

\end{pspicture}

\caption{The construction of the graph $G$ in the proof of Lemma~\ref{lem:dual}.
The edges relating to just one generic triple $(w_i,x_j,y_k)\in M$ are shown.
``Link edges'' are dashed.}
\label{fig:G}

\end{figure}

The vertex set of~$G$ (refer to Figure~\ref{fig:G}) is
$$
V=\{r\}\cup W\cup X\cup Y\cup\big\{a_{ijk},\abar_{ijk}:
   (w_{i},x_{j},y_{k})\in M\big\},
$$
and the edge set $E=T\cup L$ where
\begin{align*}
T&=\big\{\{r,w_{i}\},\{r,x_{i}\},\{r,y_{i}\}:1\leq i\leq n\big\}\\
   &\qquad\null\cup\big\{\{w_{i},a_{ijk}\},\{w_{i},\abar_{ijk}\}:
   (w_{i},x_{j},y_{k})\in M\big\}\\
\noalign{\noindent is the set of ``tree edges'' and}
L&=\big\{\{x_{j},a_{ijk}\},\{a_{ijk},\abar_{ijk}\},\{\abar_{ijk},y_{k}\}:
   (w_{i},x_{j},y_{k})\in M\big\}.
\end{align*}
the ``link edges''.  Observe that
$(V,T)$ is a tree, and that edges in~$L$ join leaves in the tree.
For $e\in E$, assign weight $w(e)=\beta_2$ if $e\in T$ is a tree edge,
and $w(e)=\beta_1$ if $e\in L$ is a link edge.

We're interested in evaluating $Z(G;q,w)$:
\begin{align}
    Z(G;q,w)&=\sum_{A\subseteq E}w(A)q^{\kappa(A)}\notag\\
    &=\sum_{B\subseteq L}\sum_{C\subseteq T}
       w(B\cup C)q^{\kappa(B\cup C)}\notag\\
    &=\sum_{B\subseteq L}h(B,\beta_2)\,\beta_1^{|B|},\label{eq:ZinH}\\
\noalign{\noindent where}
    h(B;\beta_2)&=\sum_{C\subseteq T}\beta_2^{|C|}q^{\kappa(B\cup C)}
    =Z(G\setminus\Bbar/B;q,w)\label{eq:hExpand}
\end{align}
Here, $G\setminus\Bbar/B$ denotes the graph obtained from~$G$
by deleting edges in $\Bbar=L\setminus B$ and contracting
edges in~$B$.

Let $m=|M|$, and note that $|V|=3n+2m+1$, $|T|=3n+2m$ and $|L|=3m$.
Our calculation of $Z(G;q,w)$ is greatly simplified if we take
$\beta_2$ to be exactly $-q$, rather than merely a close approximation.
So let's first determine, as a function of~$\delta$, the absolute error
we would introduce by replacing $\beta_2$ by~$-q$.
Denote by $\wmod:E\to\mathbb{Q}$ the weight function
$$
\wmod(e)=\begin{cases}
   -q,&\text{if $e\in T$;}\\
   w(e)=\beta_1,&\text{otherwise.}
\end{cases}
$$
We wish to estimate the absolute error $\bigl|Z(G;q,w)-Z(G;q,\wmod)\bigr|$.
Set $\qbar=\max\{|q|,1\}$;  then
either $|q|/|\beta_2|\geq 1$, in which case $|\beta_2| \leq \qbar$
or $|q|/|\beta_2|<1$. In this case,
since $|1+q/\beta_2|\leq 1/2$,
$|1+q/\beta_2|= 1 - |q|/|\beta_2| \leq 1/2$,
so $|\beta_2|\leq 2|q|$. We conclude that, in either case,
$|q|,|\beta_2| \leq 2\qbar$.
Furthermore,
for all $i\geq 1$, we have
$$
\beta_2^i-(-q)^i
   =
   (\beta_2+q)
\sum_{j=0}^{i-1}\beta_2^j {(-q)}^{i-1-j}
   \leq i(2\qbar)^{i-1}2 \qbar \delta =
  i {(2\qbar)}^i \delta,
$$
since $|\beta_2+q| \leq |\beta_2| \left| 1 + q/\beta_2 \right| \leq
2 \qbar \delta$.
Expanding $h(B,\beta_2)$ and $h(B,-q)$ according to~(\ref{eq:hExpand}),
and comparing term-by-term, we find that
\begin{align*}
|h(B,\beta_2)-h(B,-q)|
&\leq 2^{|T|}|T|\,(2\qbar)^{\,|T|}\delta\,\qbar^{\,|V|}\\
&\leq |T| {(2\qbar)}^{|V|+|T|} \delta\\
&=(3n+2m){(2\qbar)}^{6n+4m+1}\delta.
\end{align*}
So from (\ref{eq:ZinH}), recalling
$|\beta_1|\leq |q| \epsilon \leq \qbar$,

\begin{align}
\bigl|Z(G;q,w)-Z(G;q,\wmod)\bigr|
   &\leq 2^{|L|} {\qbar}^{|L|}
(3n+2m){(2\qbar)}^{6n+4m+1}\delta\notag\\
&\leq (3n+2m)(2\qbar)^{6n+7m+1}\delta.\label{eq:w:wtilde}
\end{align}
We'll chose $\delta$ later to make this estimate small enough.

We now proceed with our calculation, using $\wmod$ in place of~$w$, i.e.,
$-q$ in place of~$\beta_2$.
Partition sum (\ref{eq:ZinH}) in two pieces:
$$
Z(G;q,\wmod)=\subsum_{\leq}+\subsum_{>},
$$
where
$$
\subsum_{\leq}=\sum_{B\subseteq L: |B|\leq n+m}h(B,-q)\,\beta_1^{|B|}\quad
\text{and}\quad
\subsum_{>}=\sum_{B\subseteq L: |B|> n+m}h(B,-q)\,\beta_1^{|B|}.
$$
Set $Q=(-1)^{n}q^{2n+m+1}(q-1)^{m}(q-2)^{n}$, and
note that $Q\not=0$.
We'll show:
\begin{enumerate}
    \item  If $|B|<n+m$ then $h(B,-q)=0$.\label{firstObs}

    \item If $|B|=n+m$ then\footnote{{\it Bridge connected\/}
       is a synonym for 2-edge-connected, i.e., connected and having no bridge,
       which is an edge whose removal would disconnect the graph.}
    $$h(B,-q)=\begin{cases}
       Q,&\text{if $(V,T\cup B)$ is bridge connected};\\
       0,&\text{otherwise}.
    \end{cases}$$
    \label{2ndObs}

    \item  The set
    $\{B:\text{$|B|=n+m$ and $(V,T\cup B)$ is bridge connected}\}$
    is in 1-1 correspondence with the set of solutions to the
    instance of \#\textsc{3-d Matching}.\label{lastObs}
\end{enumerate}
Observations \ref{firstObs}--\ref{lastObs} entail
$$\subsum_{\leq}=Q N{\beta_1}^{n+m},$$
where $N$ is the number of solutions to the \#\textsc{3-d Matching}
instance.
On the other hand, $\subsum_{>}$ is crudely bounded
as follows:
\begin{align*}
|\subsum_{>}|&
= \sum_{B\subseteq L: |B|>n+m} h(B,-q) q^{|B|}
{\left(\frac{\beta_1}{q}\right)}^{|B|}\\
&\leq\sum_{B\subseteq L}\bigl|h(B,-q) \qbar^{|B|}{(\beta_1/q)}^{n+m+1}\bigr|\\
   &\leq 2^{|L|}2^{|T|}\qbar^{\,|T|}\qbar^{\,|V|}\,
\qbar^{|L|}
   |\beta_1/q|^{n+m+1}\\
   &\leq(2\qbar)^{6n+7m+1}|\beta_1/q|^{n+m+1}.
\end{align*}
Let $\hatQ = q^{n+m}Q$.
Now, setting $\varepsilon$ (the bound on~$|\beta_1/q|$) so that
$$
(2\qbar)^{6n+7m+1}\varepsilon\leq\tfrac18|\hatQ|,
$$
we have
\begin{equation}\label{eq:wtilde:N}
\left|\frac{Z(G;q,\wmod)}{Q\beta_1^{n+m}}-N\right|
   =\left|\frac{\subsum_{\leq}+\subsum_{>}}{Q\beta_1^{n+m}}-N\right|
   =\left|\frac{\subsum_{>}}{Q\beta_1^{n+m}}\right|
=\left|\frac{\subsum_{>}}{\hatQ{(\beta_1/q)}^{n+m}}\right|
   \leq \frac18.
\end{equation}
Then, according to~(\ref{eq:w:wtilde}), setting
$$
(3n+2m)
(2\qbar)^{6n+7m+1}\delta\leq \tfrac18Q\beta_1^{n+m},
$$
ensures
\begin{equation}\label{eq:w:wtilde'}
\left|\frac{Z(G;q,w)-Z(G;q,\wmod)}{Q\beta_1^{n+m}}\right|\leq\frac18.
\end{equation}
Combining (\ref{eq:wtilde:N}) and~(\ref{eq:w:wtilde'})
yields the required estimate
$$
\left|\frac{Z(G;q,w)}{Q\beta_1^{n+m}}-N\right|\leq\frac14.
$$

It remains to verify the three observations.  Suppose
the graph $H=G\setminus\Bbar/B$ contains a bridge~$e$.
Then
$$Z(H;q,\wmod) =\sum_{A' \subseteq E(H)-\{e\}}
\wmod(A')q^{\kappa(A')}\left(\frac{-q}{q}+1\right)=0,$$
where the $-q/q$ comes from including~$e$ in~$A$,
which gives a weight of $-q$ but reduces the number of components
by one and the $1$ comes from excluding~$e$ from~$A$.
The tree $(V,T)$ has $2(n+m)$ leaves, so if $|B|<n+m$
there are at least two vertices in $(V,T\cup B)$
of degree one.  The unique edge~$e$ incident at
either of these vertices is a bridge, and is a member
of~$T$;  it is clearly also a bridge in~$H$.
This deals with Observation~\ref{firstObs}.

Suppose $B\subseteq L$ is a set of link-edges of size $n+m$
such that $(V,T\cup B)$ is bridge connected.  Every
leaf of $(V,T)$ must have some edge of~$B$ incident at it,
and hence exactly one.  Call
such a $B$ a {\it pairing}.  If $B$ is a pairing then,
for every triple $(w_{i},x_{j},y_{k})\in M$,
either (i)~$\{x_{j},a_{ijk}\},\{\abar_{ijk},y_{k}\}\in B$
and $\{a_{ijk},\abar_{ijk}\}\notin B$,
or (ii)~$\{a_{ijk},\abar_{ijk}\}\in B$ and
$\{x_{j},a_{ijk}\},\{\abar_{ijk},y_{k}\}\notin B$.
Let $M'$ be the set of triples of type~(i).  By counting,
$|M'|=n$.  So there is a 1-1 correspondence between
pairings~$B$ and sets~$M'\subseteq M$ containing $n$~triples.
We will now argue that, under this correspondence,
bridge-connected graphs $(V,T\cup B)$ are associated
with solutions to \#\textsc{3-d Matching} and vice versa.

On the one hand, if $M'$ covers all of $W\cup X\cup Y$,
then it is easy to check that every edge in $(V,T\cup B)$
is contained in a simple cycle of the form
$(r,x_{j},a_{ijk},w_{i},r)$
or $(r,y_k,\abar_{ijk},w_i,r)$ for
some triple $(w_{i},x_{j},y_{k})\in M'$,
or a cycle of the form $(w_i,a_{ijk},\abar_{ijk},w_i)$
for some triple $(w_{i},x_{j},y_{k})\in M\setminus M'$,
Conversely, if $(V,T\cup B)$ is bridge connected then
in particular $B$ is a pairing, which immediately
implies that every element of $X$ and~$Y$ is covered
by some triple in~$M'$.  But also every $w_{i}$ must be
covered, since the only way to avoid $\{r,w_{i}\}$
being a bridge is to have either
$\{a_{ijk},x_{j}\}\in B$ or $\{\abar_{ijk},y_{k}\}\in B$,
for some $j,k$ (and hence, in fact, both).
This is Observation~\ref{lastObs}.

Finally to Observation~\ref{2ndObs}.
If $(V,T\cup B)$ is not bridge
connected then it has a bridge~$e$ which is necessarily
a tree edge.  (The link edges join leaves of the tree $(V,T)$,
and hence cannot be bridges.)  We have already seen that
the existence of a bridge implies $h(B,-q)=0$.
So suppose $(V,T\cup B)$ is bridge connected, and let
$M'$ be the corresponding 3-d matching.
Then graph $H=G\setminus\Bbar/B$ may be described as follows.

For each triple $t=(w_{i},y_{j},z_{k})\in M'$,
denote by $H_{t}=(V_{t},E_{t})$ the graph with vertex set
$$
V_{t}=\{r,w_{i},x_{j},y_{k}\}
\cup\big\{a_{ij'k'}:(w_{i},x_{j'},y_{k'})\in M\setminus M'\big\}
$$
and edge set
\begin{align*}
E_{t}
&=\big\{\{r,w_{i}\},\{r,x_{j}\},\{r,y_{k}\},
   \{w_{i},x_{j}\},\{w_{i},y_{k}\}\big\}\\
&\qquad\null\cup\big\{\{w_{i},a_{ij'k'}\}:
   (w_{i},x_{j'},y_{k'})\in M\setminus M'\big\}.
\end{align*}
The edges with endpoints of the form $a_{ij'k'}$ have multiplicity
two, the others multiplicity one.
Observe that $V_{t}\cap V_{t'}=\{r\}$
for distinct triples $t'\not=t$.
(This is a consequence of $M'$ being a 3-d matching.)
The graph $H$ is obtained by taking the union of all
$H_{t}$ and identifying vertex~$r$, so
$Z(H;q,\wmod)=q^{1-n}\prod_{t}Z(H_{t};q,\wmod)$.
Each of the multiplicity-two edges ($m-n$ of them)
contributes a factor $q(q-1)$, which is non-zero by assumption.
That leaves us with $n$~copies of $K_{4}$ minus an edge.
Each of those contributes a factor
$-q^{4}(q-1)(q-2)$, which again is non-zero, by assumption.
Putting it all together,
\begin{align*}
    Z(H;q,\wmod)&=q^{1-n}[-q^{4}(q-1)(q-2)]^{n}[q(q-1)]^{m-n}\\
    &=(-1)^{n}q^{2n+m+1}(q-1)^{m}(q-2)^{n}.
\end{align*}

Finally, we need to relate our conveniently chosen weights,
$\beta_1$ and~$\beta_2$, to the actual ones, $\alpha_1$
and~$\alpha_2$.  This is done as in the proof of Lemma~\ref{lem:bedrock}.
In particular, we choose $k_1$ and $k_2$
satisfying
${(q/\alpha_1+1)}^{k_1}-1\geq 1/\epsilon$
and
$|{(q/\alpha_2+1)}^{k_2}|<\delta$.
Now $G'$ is formed as in the proof of Lemma~\ref{lem:bedrock}
except that $k$-stretches are used in place of $k$-thickenings
(according to Equation~\ref{eq:stretchThickenRules'}).
As before, $k_1=O(m)$ and $k_2=O(m^2)$,
so the construction is polynomially bounded.
\end{proof}

Using Lemma~\ref{lem:dual}, we can now prove
Theorem~\ref{thm:shift'}.

\noindent{\bf Theorem \ref{thm:shift'}}
{\sl
Suppose $(x,y)\in \mathbb{Q}^2$ satisfies
$q=(x-1)(y-1)\notin\{0,1,2\}$.  Suppose also that
it is possible to shift the point
$(x,y)$ to the point $(x',y')$ with $x'\notin[-1,1]$,
and to $(x'',y'')$ with $x''\in(-1,1)$.
Then there is no FPRAS for
$\tutte(x,y)$
unless
$\mathrm{RP}=\mathrm{NP}$.
}

\begin{proof}
The proof is analogous to the
proof of Theorem~\ref{thm:shift}.
Note that none of $y$, $y'$ and $y''$ are equal to~$1$
since $q\neq 0$.
Let $\alpha=y-1$, $\alpha_1=y'-1$ and $\alpha_2=y''-1$.
The constraints on $x'$ and $x''$, together with
$(x'-1)(y'-1)=q$ and $(x''-1)(y''-1)=q$ imply
that $x'-1=q/\alpha_1\notin[-2,0]$ and $x''-1=q/\alpha_2\in(-2,0)$.
The proof is now exactly the same as the proof of Theorem~\ref{thm:shift}
except that Lemma \ref{lem:dual} is used in place of
Lemma~\ref{lem:bedrock}.
\end{proof}

\subsection{The hyperbola $H_2$ in the halfplane $y<-1$}

The following lemma will be used in the proof of
Lemma~\ref{lem:matchings}.

\begin{lemma}\label{lem:H2}
Suppose $\alpha_1,\alpha_2\in\mathbb{Q}-\{0\}$
satisfy $2/\alpha_1\notin[-2,0]$
and $2/\alpha_2\in(-2,0)$.
Then
$\PM\APred
\multitutte(2;\alpha_{1},\alpha_{2})$
\end{lemma}

\begin{proof}[Proof sketch.]
The construction and proof of correctness are simplified
versions of those from the proof of Lemma~\ref{lem:dual},
so we provide only a sketch here.

Suppose $G=(V,E)$ is an instance of \#\textsc{Perfect Matchings}.
For convenience, set $n=|V|/2$.
Let $\Ghat=(\Vhat,\Ehat)$ be the graph with vertex set
$\Vhat=V\cup\{t\}$ and edge set $\Ehat=E\cup T$, where
$T=\big\{\{t,v\}:v\in V\big\}$.  Define $w:\Ehat\to\{\beta_1,\beta_2\}$
by $w(e)=\beta_1$ if $e\in E$ and $w(e)=\beta_2$ in $e\in T$.
As before, $\beta_1/q$ is small in absolute value, and $\beta_2$ is
close to $-q=-2$; specifically, $|\beta_1/q|\leq\varepsilon$ and
$|1+2/\beta_2|\leq\delta$.

Following the now familiar path,
\begin{align}
    Z(G;2,w)&=\sum_{A\subseteq \Ehat}w(A)2^{\kappa(A)}\notag\\
    &=\sum_{B\subseteq E}\sum_{C\subseteq T}
       w(B\cup C)2^{\kappa(B\cup C)}\notag\\
    &=\sum_{B\subseteq E}h(B,\beta_2)\,\beta_1^{|B|},\notag\\
\noalign{\noindent where}
    h(B;\beta_2)&=\sum_{C\subseteq T}\beta_2^{|C|}2^{\kappa(B\cup C)}
\end{align}
Set $Q=q^{n+1}(q-1)^n=2^{n+1}$.  We will
show the following observations.
\begin{enumerate}
    \item  If $|B|<n$ then $h(B,-2)=0$.

    \item If $|B|=n$ then
    $$h(B,-2)=\begin{cases}
       Q,&\text{if $(\Vhat,T\cup B)$ is bridge connected};\\
       0,&\text{otherwise}.
    \end{cases}$$

    \item  The set
    $\{B:\text{$|B|=n$ and $(\Vhat,T\cup B)$ is bridge connected}\}$
    is in 1-1 correspondence with the set of solutions to the
    instance of \#\textsc{Perfect Matchings}.  Specifically,
    $(\Vhat,T\cup B)$ is bridge connected iff $B$ is a perfect
    matching in~$G$.
\end{enumerate}
Thus, for $\varepsilon,\delta$ sufficiently small,
$\bigl|2^{-(n+1)}\beta_{1}^{-n}Z(G;q,w)-N\bigr|\leq\tfrac14$,
where $N$ is the number of perfect matchings in~$G$.
The proof is completed exactly as before.

It remains to justify the three observations.
For Observation~1, note that if $|B|<n$ then $(V,B)$ contains
an isolated vertex. Consider the factor
contributed to $h(B,-2)$ from the edge
connecting this vertex to~$t$.
The contribution is $-q$ (for including the edge)
plus $q$ (for excluding it, and hence adding a component),
which is~$0$.
Observation~3 is self-evident.
Using Observation~3, we can establish
Observation~2 as follows.
Suppose that $B$ is a perfect matching.
Then
$$h(B,-2) = 2{\left({(-2)}^2+2(-2)+2\right)}^n=2^{n+1},$$
where the first $2$ comes from the component containing~$t$,
and for each of the $n$ edges in the matching,
the ${(-2)}^2$ comes from including both edges to~$t$,
the $2(-2)$ comes from the two ways to add one of
the edges to~$t$, and the $2$ comes from excluding
both edges to~$t$, which adds a component.

\end{proof}

We can now prove Lemma~\ref{lem:matchings}.

\noindent {\bf Lemma
\ref{lem:matchings}.\quad}{\sl
Suppose $(x,y)$ is a point with $(x-1)(y-1)=2$ and $y<-1$.
Then $\tutte(x,y) \APequiv \PM$.
}

\begin{proof}

To show
$\PM \APred \tutte(x,y)$
use thickenings as in the proof of Corollary~\ref{cor:y<-1}
to shift $(x,y)$ to a point $(x',y')$ with $x'\notin[-1,1]$
and to a point $(x'',y'')$ with $x''\in(-1,1)$. Then follow the
proof of Theorem~\ref{thm:shift'}
to reduce $\multitutte(2;\alpha_1,\alpha_2)$ to
$\tutte(x,y)$. Finally,
Lemma~\ref{lem:H2} reduces $\PM$ to $\multitutte(2;\alpha_1,\alpha_2)$.

We now show $\tutte(x,y)\APred \PM$.
Using the Fortuin-Kasteleyn representation of the Potts model,
$Z(G;2,y-1)$ is equal to the partition function of the
Ising model in which every edge has weight $y-1$.
That is,
\begin{equation}
\label{eq:Potts}
Z(G;2,y-1)=
\sum_{\sigma:V(G)\rightarrow\{-1,1\}}
y^{\mathrm{mono}(\sigma)},
\end{equation}
where $\mathrm{mono}(\sigma)$ denotes the number of monochromatic
edges in the mapping~$\sigma$.
(See \cite[(2.7), (2.9)]{Sokal05} for a justification of this identity.)
We will assume without loss of generality that the graph $G$ has no loops. It
is clear from (\ref{eq:Potts}) that a loop merely introduces a factor of~$y$.

Let $n=|V(G)|$ and $m=|E(G)|$. Let $\nu=(y-1)/(y+1)$.
Now let $G'$ be the graph constructed from $G$
by replacing each vertex of degree $\ell\geq 4$
as follows. Suppose that the neighbours of vertex $v$ in~$G$
are $w_1,\ldots,w_\ell$.
Then replace $v$ with a path of $\ell-2$ new degree-$3$ vertices
$v_2,\ldots,v_{\ell-1}$.
The edges $(v_2,v_3),(v_3,v_4),\ldots,(v_{\ell-2},v_{\ell-1})$
will be called ``supplementary'' edges of $G'$.
The edges $(w_1,v),(w_2,v),\ldots,(w_{\ell-1},v),(w_\ell,v)$
correspond to edges
$(w_1,v_2),(w_2,v_2),(w_3,v_3),\ldots,(w_{\ell-1},v_{\ell-1}),(w_\ell,v_{\ell-1})$
of $G'$.
We will call these edges ``primary'', because they correspond to the original
edges of~$G$.

Then let $G''$ be the graph constructed from $G'$ by replacing
each vertex of degree $2$ as follows. Suppose that the neighbours of vertex
$v$ in $G'$ are $w_1$ and $w_2$. Then replace $v$ with
two new vertices $v_1$ and $v_2$
and replace the edges $(w_1,v)$ and $(w_2,v)$ with the path
$(w_1,v_1),(v_1,v_2),(v_2,w_2)$ in which $(v_1,v_2)$ is a supplementary
edge of $G''$ and
the edges $(w_1,v_1)$ and $(v_2,w_2)$ are ``primary'' edges of $G''$.
Finally, if $v$ is a degree-$3$ vertex of $G'$ with neighbours $w_1$,
$w_2$ and $w_3$,
replace $v$ with the three vertices $v_1,v_2,v_3$.
Add supplementary edges $(v_1,v_2),(v_2,v_3),(v_3,v_1)$.
Replace the edges $(w_1,v)$, $(w_2,v)$ and $(w_3,v)$ of $G'$ with
edge $(w_1,v_1)$, $(w_2,v_2)$ and $(w_3,v_3)$ of $G''$, making edge
$(w_i,v_i)$ primary if and only if $(w_i,v)$ was primary in $G'$.
(Once again, the primary edges of $G''$ correspond
to the original edges of~$G$.)

Fisher has shown \cite[(10)]{Fisher} that
\begin{equation}
\label{eq:fisher}
Z(G;2,y-1)=y^m 2^n {\left(\frac{\nu}{1+\nu}\right)}^m
\sum_{X} \prod_e \frac{1}{\nu},
\end{equation}
where the sum is over perfect matchings $X$ of $G''$
and the product is over primary edges $e$ of $G''$ that are in the
perfect matching~$X$.

Now let $n_1$ and $n_2$ be positive integers such that
$1/\nu=n_1/n_2$.
Let $H$ be a graph consisting of $n_1$ parallel edges
from a vertex~$u$ to a vertex~$a$ and $n_2$ parallel edges
from the vertex~$a$ to a vertex~$b$ and a single edge from~$b$ to
a vertex~$v$.
Let $\mathcal{M}$ be the set of matchings of $H$ which match
both~$a$ and~$b$.
There are $n_1$ matchings in $\mathcal{M}$ in which $a$ is matched
with~$u$. All of these match both~$u$
and~$v$.
There are $n_2$ matchings in $\mathcal{M}$ in which $a$ is matched
with~$b$. These do not match~$u$ or~$v$.
There are no other matchings in~$\mathcal{M}$.

Construct $\hatG$ from $G''$ by replacing every primary edge $(u,v)$
of~$G''$ with a copy of~$H$.
Then the expression $\sum_{X} \prod_e \frac{1}{\nu}$ in
Equation~(\ref{eq:fisher}) is equal to the number of perfect matchings of
$\hatG$ divided by~$n_2^m$.
So if we could approximate the number of perfect matchings of $\hatG$,
we could approximate $Z(G;2,y-1)$.

\end{proof}

\noindent{\bf Remark.}
The construction in
the reduction from the problem $\tutte(x,y)$ to
the problem $\PM$ relies on the
fact that $y-1$ and $y+1$ have the same sign (so $n_1$ and $n_2$
are both positive integers).
The same reduction would apply for $q=2$ and $y>1$
but this is ferromagnetic Ising, and we already have an FPRAS,
due to Jerrum and Sinclair~\cite{JS93}.

\section{\#P-hardness}
\label{sec:nump}

In Section~\ref{sec:innumP},
we noted that
if $x\geq 0$ and $y\geq 0$ then $\tutte(x,y)$ is in
$\numPQ$, so
there is a randomised approximation scheme for $\tutte(x,y)$ that
runs in polynomial time using an oracle for an NP predicate.
Here we show that it is unlikely that $\tutte(x,y)$
is in $\numPQ$ for all $x$ and~$y$.
In particular, we identify a region of points $(x,y)$ where
$y$ is negative for which even approximating $\tutte(x,y)$
is as hard as~\#P.
Specifically, we prove the following.

\begin{theorem}
\label{thm:nump}
Suppose $(x,y)$ is a point with $y\in (-1,0)$
and $(x-1)(y-1)=4$. Then there is no FPRAS for $\tutte(x,y)$
unless $\mathrm{RP}=\mathrm{\#P}$.
\end{theorem}

\subsection{The Potts model}

For a positive integer $q$ and a $y\in\mathbb{Q}$,
and a graph $G=(V,E)$,
let $$P(G;q,y) =
\sum_{\sigma:V\rightarrow\{1,\ldots,q\}}
y^{\mathrm{mono}(\sigma)},$$
where $\mathrm{mono}(\sigma)$ is the number of edges in~$E$
that are monochromatic under the map~$\sigma$.
$P(G;q,y)$ is the partition function of the {\it $q$-state Potts model\/}
at an appropriate temperature (depending on $y$).
The region $y\geq 1$ is ``ferromagnetic'' since like spins are
favoured along an edge, the region $0\leq y \leq 1$ is
``antiferromagnetic'', and the region $y\leq 0$ is
``unphysical'' \cite{Sokal05}.
It is known that the $q$-state Potts model coincides with the Tutte
polynomial when $q$ is a positive integer.
In particular (see (\ref{eq:rcequiv}) and
\cite[(2.9)]{Sokal05}),
$$ T(G;x,y) = {(y-1)}^{-n}{(x-1)}^{-\kappa(E)} P(G;q,y),$$
where $q=(x-1)(y-1)$.

In the rest of this section, we
suppose that we have an FPRAS for $P(G;4,y)$
for a point $y \in (-1,0)$ and we show how to
use the FPRAS to solve a \#P-hard problem
(counting proper $3$-colourings of a simple graph).

First, we establish some notation.
If $G$ is a graph with designated vertices $a$ and $b$ and
$\alpha$ and $\beta$ are values in $\{1,\ldots,q\}$,
let $P(G;q,y \mid \sigma(ab)=\alpha\beta)$ denote the
contribution to $P(G;q,y)$ due to colourings~$\sigma$
with $\sigma(a)=\alpha$ and $\sigma(b)=\beta$.

\subsection{The building blocks}
\label{sec:buildingblocks}

Fix $y\in (-1,0)$.
Suppose that $n$ is the number of vertices of a graph $G$.
Let $M$ be a
rational number in the range $1\leq M \leq 3^n$ and let
$\epsilon=2^{-n^2}$.
In this section,
we will show how to construct a graph $H_M$ with
two designated vertices, $a$ and $b$, so that
\begin{equation}
\label{eq:gadget}
\frac{-1}{M} \leq
\frac{P(H_M;4,y \mid \sigma(ab)=11)}{P(H_M;4,y \mid \sigma(ab)=12)}
\leq \frac{-1}{M} + \epsilon
 .\end{equation}

As a building block, let $P_{\ell}$ be an $\ell$-edge path.
Let $f_\ell$ denote $P(P_\ell;4,y \mid \sigma(ab)=11)$
and let $a_\ell$ denote $P(P_\ell;4,y \mid \sigma(ab)=12)$.
These satisfy the recurrences
$f_\ell = y f_{\ell-1} + 3 a_{\ell-1}$ and
$a_\ell = f_{\ell-1} + (2+y) a_{\ell-1}$ with $f_1=y$ and $a_1=1$.
The solution to these recurrences, for $\ell\geq 1$,
is given by
$$ f_\ell = \frac14 {(3+y)}^{\ell}+ \frac34 {(y-1)}^{\ell},$$
and
$$a_\ell = \frac14 {(3+y)}^{\ell}-\frac14 {(y-1)}^{\ell}.$$
Thus,
$$\frac{a_\ell}{f_\ell} =
\frac{(3+y)^\ell - (y-1)^\ell}
{(3+y)^\ell + 3 (y-1)^\ell} =
1 -
\frac{4(y-1)^\ell}{(3+y)^\ell+ 3(y-1)^\ell}.$$
Recall $y>-1$ and let
$\gamma={\left((3+y)/(y-1)\right)}^2>1$.
For every positive integer $j$, let
$\delta_j = 1-a_{2j}/f_{2j}$.
Note that
\begin{equation}\label{eq:gamma}
0<\gamma^{-j}<\delta_j < 4 \gamma^{-j}.\end{equation}
Also, $f_{2j}/a_{2j}=1/(1-\delta_j)$.

Given $y$, choose a positive odd integer $k$ so that
\begin{equation}
\label{eq:delta}
|y|^k \leq \frac{1}{M}
< |y|^{k-2}.
\end{equation}
Now, let $t$ be the smallest integer such that $\delta_t \leq
\epsilon M$.
For $j\in\{1,\ldots,t\}$,
choose a natural number $k_j$ so that
\begin{equation}
\label{eq:nump}
|y|^k
\prod_{r=1}^{j-1}
\frac{1}{(1-\delta_r)^{k_r}}
\frac{1}{(1-\delta_j)^{k_j}}
\leq \frac{1}{M}
< |y|^k
\prod_{r=1}^{j-1}
\frac{1}{(1-\delta_r)^{k_r}}
\frac{1}{(1-\delta_j)^{k_j+1}}
.\end{equation}

Now $H_M$ is formed by joining a number of paths, all with
endpoints~$a$ and~$b$.
To form $H_M$, take $k$ paths of length~$1$ (i.e., edges).
Also, for every $j\in\{1,\ldots,t\}$, take $k_j$
paths of length $2j$.
So
$$\frac{P(H_M;4,y\mid\sigma(ab)=11)}{P(H_M;4,y\mid\sigma(ab)=12)}
=- |y|^k
\prod_{r=1}^{t}
\frac{1}{(1-\delta_r)^{k_r}},$$
and this is at least $-1/M$ and at most $-(1/M)(1-\delta_t)$,
which implies Equation~(\ref{eq:gadget}).

Now we consider the size of $H_M$.
Equation~(\ref{eq:delta}) implies
that $k=O(\log M) = O(n)$.  Also, $t=O(n^2)$
by (\ref{eq:gamma}).

How big can $k_j$ be?
By (\ref{eq:nump}) we have
$
\frac{1}{(1-\delta_j)^{k_j}}\leq \frac{1}{1-\delta_{j-1}}$,
so ${(1-\delta_j)}^{k_j} \geq 1-\delta_{j-1}$.
$\delta_j$ is decreasing in $j$, so without loss of generality,
we'll deal with those $j$ such that $\delta_j\leq 0.7$
(the values of $k_j$ corresponding to smaller values of $j$
are just constants).
Then
$${(1-\delta_j)}^{2/\delta_j} < 0.15 <
{(1-\delta_{j-1})}^{1/\delta_{j-1}},$$
so
$$
{(1-\delta_j)}^{2 \delta_{j-1}/\delta_j} <
{(1-\delta_{j-1})}^{\delta_{j-1}/\delta_{j-1}} = 1-\delta_{j-1},$$
and therefore $k_j \leq 2 \delta_{j-1}/\delta_j = O(1)$.

\subsection{The construction}

We use the notation from Section~\ref{sec:buildingblocks}.
Let $r$ be the smallest even integer such that
$|y|^r<\epsilon 4^{-n}$.
Construct $G'$ from the simple graph $G$ (the graph we wish to 3-colour)
by replacing every edge $(u,v)$ of~$G$
with a bundle of $r$~parallel edges with endpoints~$u$ and~$v$.
(That is, we perform an $r$-thickening on all edges.)
Add two new vertices, $a$ and~$b$.
Connect $a$ to every vertex in~$G$ by a bundle of $r$~parallel edges.
Similarly, connect $b$ to every vertex in~$G$.

Let $n$ denote the number of vertices of $G$ and $m$ denote the
number of edges of $G$.
Recall that $P(G;3,0)$ is the number of proper $3$-colourings of~$G$.
Then,
$$P(G;3,0) \leq{P(G';4,y \mid\sigma(ab)=11)}\leq P(G;3,0) + 4^n |y|^r,$$
so
\begin{equation}
\label{eq:useful1}
P(G;3,0) \leq P(G';4,y \mid \sigma(ab)=11)\leq P(G;3,0)  + \epsilon.
\end{equation}
Similarly,
\begin{equation}
\label{eq:useful2}
P(G;2,0) \leq P(G';4,y \mid \sigma(ab)=12)\leq P(G;2,0) +\epsilon.
\end{equation}

Let $G_M$ be the graph constructed from $G'$ and $H_M$
by identifying vertex $a$ in $G'$ with vertex $a$ in $H_M$
and similarly identifying vertex $b$ in $G'$ with vertex $b$ in $H_M$.
Let
$$Y_M =\frac{P(G_M;4,y)}{P(H_{M};4,y \mid \sigma(ab)=12)}$$
and note (from the previous section) that the
quantity $P(H_{M};4,y \mid\sigma(ab)=12)$ in the denominator
is positive.
Now
$$P(G_M;4,y \mid\sigma(ab)=11) =
P(H_{M};4,y \mid\sigma(ab)=11) P(G';4,y \mid \sigma(ab)=11),$$
and
$$P(G_M;4,y) = 4 P(G_M;4,y \mid\sigma(ab)=11)
   + 12 P(G_M;4,y \mid\sigma(ab)=12),$$
so
\begin{align*}
Y_M &= 4\,\frac{P(H_{M};4,y \mid\sigma(ab)=11)}
{P(H_{M};4,y \mid\sigma(ab)=12)}\,P(G';4,y \mid\sigma(ab)=11)\\
 &\qquad\null+ 12P(G';4,y \mid\sigma(ab)=12).
\end{align*}
Let $\xi=5\epsilon 3^n = o(1)$.
By Equations~(\ref{eq:gadget}), (\ref{eq:useful1}), and
(\ref{eq:useful2}),
$$
Y_M = - 4\,
\frac{P(G;3,0)}{M}
+ 12 P(G;2,0) + \rho_M,
$$
where $|\rho_M|\leq \xi$.

We will restrict attention to graphs $G$ which are bipartite
with at least $4$~vertices.
Note that it is  \#P-hard to count the proper
$3$-colourings of a bipartite graph. For example,
\cite[Section 6]{DGGJ04} observes that
this is the
same as counting homomorphisms from a general graph to
the cycle $C_6$, which is shown to be \#P-hard by Dyer
and Greenhill~\cite{DyerGreenhill}.
Also, for such a graph $G$, $P(G;2,0)>0$ and
$P(G;3,0)\geq 4 P(G;2,0)$.

Now, suppose that we had an FPRAS for
approximating $P(G_M;4,y)$.
A call to the FPRAS gives us the sign of $Y_M$.

Let $G$ be a bipartite graph with $n\geq4$ vertices.
Let $z_\ell = 3^{-n}$ and $z_u=1$.
Then we have an interval $[z_\ell,z_u]$
with $Y_{1/z_\ell}>0$ and $Y_{1/z_u}<0$.
Use binary search to bisect the interval
until we have
an interval $[z_\ell,z_u]$ with
$Y_{1/z_\ell}\geq 0$, $Y_{1/z_u}\leq 0$, and $z_u-z_\ell\leq \epsilon$.
(This takes at most $n^2$ bisections since, after $j$~bisections,
$z_u-z_\ell \leq 2^{-j}$.)

Since $Y_{1/z_\ell}\geq 0$, we have
$$z_\ell \leq
\frac{3 P(G;2,0)}{P(G;3,0)} + \frac{\xi}{4 P(G;3,0)}.$$
Similarly, since $Y_{1/z_u}\leq 0$,
$$z_u \geq
\frac{3 P(G;2,0)}{P(G;3,0)} - \frac{\xi}{4 P(G;3,0)}.$$
So
$$
\frac{3 P(G;2,0)}{P(G;3,0)} - \frac{\xi}{4 P(G;3,0)}
\leq z_u \leq z_\ell+\epsilon \leq
\frac{3 P(G;2,0)}{P(G;3,0)} + \frac{\xi}{4 P(G;3,0)} + \epsilon.$$

Now the point is that only one real number in the
specified interval for $z_u$ is of the form $3 n_1 / n_2$
where $n_1$ is an integer in $\{1,\ldots,2^n\}$ and
$n_2$ is an integer in $\{1,\ldots,3^n\}$
(since $\epsilon$ and $\xi$ are so small)  so
the value of $z_u$ allows us to compute
$3 P(G;2,0)/P(G;3,0)$ exactly, and since $P(G;2,0)$ can be computed exactly,
this gives us $P(G;3,0)$, thus counting proper $3$-colourings of~$G$.

\bibliographystyle{plain}
\bibliography{TutteApprox}

\end{document}